\documentclass[12pt]{article}

\let\tilde=\widetilde
\let\hat=\widehat
\newcommand{\beqa}{\begin{eqnarray}}
\newcommand{\eeqa}{\end{eqnarray}}
\newcommand{\beq}{\begin{equation}}
\newcommand{\eeq}{\end{equation}}
\newcommand{\rec}[1]{\mbox{$\frac{1}{#1}$}}
\newcommand{\mfrac}[2]{\mbox{$\frac{#1}{#2}$}}
\newcommand{\half}{\mbox{$\frac{1}{2}$}}
\newcommand{\nn}{\nonumber}

\newcommand{\de}{\delta}
\newcommand{\etal}{{\em et al.}}

\newcommand{\ep}{\varepsilon}
\newcommand{\ee}{{\rm e}}
\newcommand{\ol}[1]{\overline{#1}}
\newfont{\bbold}{msbm10 scaled\magstep1}

\newcommand{\lbl}[1]{\label{#1}}

\begin{document}

\title{Symmetry-breaking in chiral polymerisation}
\author{Jonathan AD Wattis$^*$ and 
Peter V Coveney$^\dag$ \\ 
$^*$Theoretical Mechanics, 
School of Mathematical Sciences, \\ 
University of Nottingham, University Park, 
Nottingham NG7 2RD, U.K.\\  
Jonathan.Wattis@nottingham.ac.uk \\ 
Tel:  +44 (0)115 951 3857 Fax:  +44 (0)115 951 3837 \\ 
$^\dag$Centre for Computational Science, Department of Chemistry, \\ 
University College London, 20 Gordon Street, London, WC1H 0AJ, UK.\\ 
P.V.Coveney@ucl.ac.uk\\
Tel: +44 (0)20 7679 4560 Fax: +44 (0)20 7679 7463 }
\date{17 February 2004}

\maketitle

%-------------------------------------------------------

\begin{abstract}
We propose a model for chiral polymerisation and investigate 
its symmetric and asymmetric solutions. The model has a source 
species which decays into left- and right-handed types of monomer, 
each of which can polymerise to form homochiral chains; 
these chains are susceptible to `poisoning' by the opposite handed 
monomer.  Homochiral polymers are assumed to influence the 
proportion of each type of monomer formed from the precursor.  
We show that for certain parameter values a positive feedback 
mechanism makes the symmetric steady-state solution unstable. 

The kinetics of polymer formation are then analysed in the case 
where the system starts from zero concentrations of monomer 
and chains.  We show that following a long induction time, extremely 
large concentrations of polymers are 
formed for a short time, during this time an asymmetry 
introduced into the system by a random external perturbation 
may be massively amplified. The system then approaches one 
of the steady-state solutions described above. 
\end{abstract}
{\bf Keywords} 
autocatalysis, 
bifurcations, 
chiral polymerisation, 
cross-catalysis,
growth kinetics, 
symmetry-breaking.

%-------------------------------------------------------
\section{Introduction}

Studies of the origins of life raise many associated fundamental questions.   
Among these, one is concerned with the origin and propagation of 
molecular handedness.  
It is well known that chirality is a signature of life as we know it. 
Nucleic acids contain only D-sugars while proteins are made only from 
L-amino acids (although D-amino acides do occur in Nature and even 
occasionally show up in some proteins (Jung, 1992).  
What leads to the synthesis of {\em homochiral} polymers, 
in which all the constituent monomers have the same handedness ? 
And what is responsible for the evolution of chiral {\em purity}, 
the more or less exclusive dominance of one macromolecular handedness 
over its mirror image? These are questions of great interest and
importance and remain the subject of much discussion. 

It is known that, in general, the addition of the correct enantiomer 
to a growing polymer chain is more favourable that the wrong one 
(Joshi \etal\ 2000). Indeed Joyce \etal (1984) showed that 
addition of the wrong-handed monomer to a growing oligonucleotide 
chain acts as a chain terminator, stopping all further reaction. 
For the case of proteins, there is also the driving force of the beneficial
secondary structures, such as $\alpha$-helices and $\beta$-sheets, 
that may arise from homochiral polymers.  Given these assumptions, 
that wrong-handed monomers inhibit chain growth, our paper is 
concerned with whether, starting from a racemic mixture of monomers, 
it is possible to produce a system of homochiral polymers of a 
greater or lesser degree of chiral purity. 
Starting from an achiral substrate, we shall be concerned with whether it 
is possible to produce a system of homochiral polymers of high chiral 
purity by analysing some plausible kinetic models. 

There is some discussion of related matters and experimental observations 
in the recent literature. Zubay (2000) provides a readable 
discussion of possible pre-biotic chemistry, while Colonna \etal\
(1994) describes a number of alternative self-reproducing
systems. 
Sandars (2002) reviews the range and importance 
of chirality in biological systems, as well as the chemical processes 
which lead to achiral states.  After summarising a possible historical order of events 
in the origin of life on Earth Sandars discusses the stages where a bifurcation 
to a chiral state may occur.  He then applies  existing knowledge of 
chirality on Earth to speculate on the question of extra-terrestrial life 
and its chirality. 

Luisi's group at ETH in Zurich has studied various polymeric systems in which 
left- and right-handed monomers aggregate together to form larger than 
expected concentrations of homochiral polymers (Blocher \etal\ 2001, and 
Hitz \etal 2001). 
Hitz \etal\ present an analysis of the data and postulate that the 
exess of homochiral polymers is due to a high-order Markov process 
rather than the feedback mechanism which we analyse in this paper.  
It is readily understood that the rate coefficient governing polymer 
growth may depend on the handedness of the monomer 
to be attached to the chain and handedness of the monomer which 
currently terminates the chain; however, with a high-order Markov 
processes this rate coefficient may also depend on the  handedness 
of the penultimate (and possibly the antepenultimate) monomers in the 
polymeric chain. 

While oligopeptides spontaneously form homochiral sequences,
Hitz \& Luisi (2002) have shown that the presence of quartz 
promotes the production of a high yield of homochiral sequences. 
More recently (Hitz \& Luisi, 2003) this has been quantified and the level  
of enantiomeric exess in a system has been amplified from 20\% to 
over 70\% and in some cases to 100\%, by the presence of quartz. 

The model of symmetry-breaking in chiral polymerisation which we 
explore in this paper is based on a model suggested by Sandars 
(2003). Sandars gives an account of the 
history of chemical discoveries leading up to the mechanisms of 
enantiomeric cross-inhibition and autocatalysis upon which his 
and our models rely.   Sandars integrates the resulting system of 
ordinary differential equations numerically to explore the parameter 
regimes where symmetry-breaking solutions exist. 
He observes a `phase-transition' type of phenomenon 
where a small change in the fidelity of the feedback mechanism 
leads to a large-scale change in the steady-state which the 
system as a whole converges to. Below a critical fidelity in the 
nonlinear feedback process the 
system approaches a symmetric state where equal amounts of 
left- and right-handed polymers coexist, whilst above the critical 
fidelity a homochiral state is approached in which one chirality of 
polymer dominates to the almost complete exclusion of the other. 

The model studied here includes the inhibition of homochiral sequences 
of long chains. This bears some similarity with our modelling of 
cement hydration (Wattis \& Coveney, 1997) in which larger clusters were susceptible to 
poisoning by another component. Here the poison is simply the 
monomer of opposite chirality, so each monomer can either act as an 
agent of {\em growth} (of polymers of the same handedness) or 
of {\em inhibition} (of polymers of the opposite handedness). 
This dual role leads to some subtle and interesting effects since 
an abundance of, say, right-handed polymers makes it unlikely for any 
left-handed polymeric sequences to form, and the majority of left-handed 
monomers produced will be consumed by inhibiting right-handed 
homochiral sequences - a `double-whammy' effect. 
This form of competition is distinct to the models of nucleation 
involving competition we have analysed previously; 
for example in Wattis (1999) and in 
Bolton \& Wattis (2004)  there is only one 
monomer which assembles to form two 
morphologies of cluster.  Competition is thus between the 
growth of one type of cluster and that of another.  
However, similar mechanisms are operative in both those examples 
and in the present paper, since both types of homochiral polymer 
sequence (left and right) are ultimately composed from, 
and hence competing for, the same source material. 

We have investigated the growth of RNA chains in an earlier paper, 
(Wattis \& Coveney, 1999) wherein we used a much more detailed and hence complicated model 
to assess the feasibility of long self-replicating RNA sequences forming within a 
realistic timescale.  While that model of RNA polymerisation had no 
precursor species, it contained four types of nucleotide monomer (A, C, G and T), 
both autocatalytic and cross-catalytic polymerisation mechanisms 
and an important hydrolysis step that recycled growing RNA 
sequences.  On the basis of plausible assumptions about the prebiotic 
soup, and by invoking a number of approximations and coarse-graining 
procedures, we were able to show that self-replicating RNA sequences 
are amplified in such mixtures, while less capable replicators are driven 
to extinction.

By comparison, in the present paper we are able to construct a model which 
is more directly tractable using standard methods of mathematical analysis. 
The paper is structured as follows.  In section \ref{mod-sec}, 
we specify the basic kinetic model, while in section \ref{ss-sec} 
we study its steady-state solutions. Section \ref{kin-sec} considers 
the time-dependent achiral solutions and investigates their kinetic stability. 
Section \ref{fe1-sec} then considers the case of perfect chiral 
symmetry-breaking. It is followed by a discussion (section \ref{disc-sec}) 
and conclusions from our work are presented in section \ref{conc-sec}.  

%---------------------------------------------------------------------------
\section{The kinetic model}
\label{mod-sec}

%-----------------------
\subsection{Microscopic modelling}

We aim to investigate spontaneous symmetry breaking in a system 
which allows both right- and left-handed chiral polymers to form. 
We assume there is some achiral source $S$ which spontaneously 
transforms into right- and left-handed monomers at a slow rate $\ep$; 
we also assume that the presence of longer chiral polymers (denoted 
$L$ or $R$) accelerates the formation of monomers of the same chirality.   
Thus we shall be concerned with the following set of coupled chemical 
reactions:  
\beq \begin{array}{ccccc}
S \stackrel{\ep}{\longrightarrow} L_1 , &&
S+ Q \stackrel{\frac{k\!(\!1\!+\!\!f)}{2}}{\longrightarrow} L_1 + Q , &&
S+ P \stackrel{\frac{k\!(\!1\!-\!\!f)}{2}}{\longrightarrow} L_1 + P , \\ 
S \stackrel{\ep}{\longrightarrow} R_1 , && 
S+ P \stackrel{\frac{k\!(\!1\!+\!\!f)}{2}}{\longrightarrow} R_1 + P , &&
S+ Q \stackrel{\frac{k\!(\!1\!-\!\!f)}{2}}{\longrightarrow} R_1 + Q , 
\end{array} \lbl{Sbreakdown} \eeq
where $L_1,R_1$ are the left and right monomer species respectively while  
$Q$ ($P$) represents some measure of the total concentrations of
left-handed (right-handed) homochiral polymers in the system.  
The precise forms of these rate processes will be specified later 
(see equation (\ref{PQdef})).  The rate coefficients of the reactions are $\ep$, $k$.  
We shall assume that $\ep\ll k$, and that $k$ depends on the length 
of the polymer $L,R$.  
The parameter $f$ in these rate coefficients is the {\em fidelity} of the 
feedback mechanism; typically this will not be perfect, that is, $f<1$ is 
likely in general.  

This model has some similarities with models we have studied in  
other areas of investigation; for example, the kinetics of micelle-formation 
in ethyl caprylate  (Coveney \& Wattis, 1996).  In this system the breakdown of caprylate 
ester  into monomer occurs spontaneously at some slow rate, 
but is massively accelerated by the presence of micelles, 
which have a catalytic role in the breakdown of the source species. 
In the present paper this mechanism is more complex 
since there are {\em two} monomers, and there is an additional fidelity 
parameter since long left- or right-handed sequences can promote 
the formation of the oppositely-handed (right/left) monomer as well as its own. 
In the caprylate system, unusual kinetic behaviour is observed 
when the system is initiated without any product present.   
The system is then effectively in a metastable state, and very little appears 
to happen for a long time. Following this induction time, 
the kinetics then proceed fairly rapidly.  We expect to see similar 
behaviour in the current model in the case $\ep\ll1$. 

Sandars' model (2003) differs from ours in that his imposes a 
maximum polymer length, typically set at five, and only polymers of this 
maximum length act catalytically in the breakdown of the source into 
monomers. We allow polymers to grow to arbitrary lengths, and chiral 
polymers of all lengths have some degree of efficacy in the autocatalytic 
feedback mechanism by which the source species decays to form chiral 
monomers.  

The monomers will be allowed to combine to form chirally pure polymers, 
denoted by $L_n$ and $R_n$ according to 
\beq
L_n + L_1 \stackrel{a}{\longrightarrow} L_{n+1} , \qquad 
R_n + R_1 \stackrel{a}{\longrightarrow} R_{n+1}  . 
\eeq
We assume that the monomer of opposite handedness may attach to 
a growing polymer and so inhibit its growth; such inhibited sequences 
will be denoted by $RL_n$ for a polymer $L_n$
which has been terminated by an $R_1$ monomer, and 
$LR_n$ for the corresponding $R_nL_1$ polymer.  We 
denote the rate of such reactions by $a\chi$.  Thus we have 
the two rate processes 
\beq
L_n + R_1 \stackrel{a\chi}{\longrightarrow} RL_n , \qquad 
R_n + L_1 \stackrel{a\chi}{\longrightarrow} LR_n  . 
\lbl{poisoning} \eeq 
However, the sequences $RL_n,LR_n$ are treated as inert 
products which have no influence on the other rate processes; 
therefore their concentrations can be ignored in the mathematical 
modelling of the chemical reactions. Following Joyce \etal\ (1984) 
we assume no further growth of these products can occur.  
The system studied by Joshi \etal\ (2000) corresponds to $0<\chi<1$; 
however, we shall consider the full range of possible $\chi>0$.

%-----------------------
\subsection{Macroscopic model}

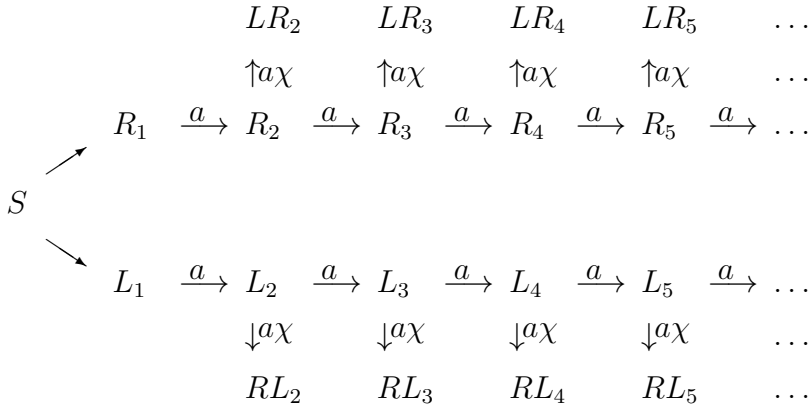
\begin{figure}[ht]
\begin{picture}(400,150)(0,-75)
\put( 50,30){$R_1$}
\put(100,30){$R_2$}
\put(150,30){$R_3$}
\put(200,30){$R_4$}
\put(250,30){$R_5$}
\put(100,70){$LR_2$}
\put(150,70){$LR_3$}
\put(200,70){$LR_4$}
\put(250,70){$LR_5$}
\put( 50,-30){$L_1$}
\put(100,-30){$L_2$}
\put(150,-30){$L_3$}
\put(200,-30){$L_4$}
\put(250,-30){$L_5$}
\put(100,-70){$RL_2$}
\put(150,-70){$RL_3$}
\put(200,-70){$RL_4$}
\put(250,-70){$RL_5$}
\multiput(75,30)(50,0){5}{$\longrightarrow$}
\multiput(79,35)(50,0){5}{$a$}
\multiput(75,-30)(50,0){5}{$\longrightarrow$}
\multiput(79,-25)(50,0){5}{$a$}
\multiput(100, 50)(50,0){4}{$\uparrow$}
\multiput(100,-50)(50,0){4}{$\downarrow$}
\multiput(105, 50)(50,0){4}{$a\chi$}
\multiput(105,-47)(50,0){4}{$a\chi$}
\multiput(300,30)(0,20){3}{$\ldots$}
\multiput(300,-30)(0,-20){3}{$\ldots$}
\put(10,0){$S$}
\put(25,15){\vector(3,2){15}}
\put(25,-10){\vector(3,-2){15}}
\end{picture}
\caption{Diagrammatic representation of the homochiral 
polymerisation scheme under study.}
\label{detail-fig}
\end{figure}

To close the system of rate equations ensuing from the scheme 
(\ref{Sbreakdown})--(\ref{poisoning}) we shall assume that the source 
chemical, $S$, is added to the system at some constant rate $S_0$. 
We define macroscopic quantities for the total concentration of all 
homochiral sequences of each 
type of polymer by $L$ and $R$, and the mass of monomers in each 
set of homochiral sequences by $Q$, $P$  in schema (\ref{Sbreakdown}) by 
\beq 
L = \sum_{n=2}^\infty L_n , \quad 
R = \sum_{n=2}^\infty R_n , \quad 
Q = \sum_{n=2}^\infty n L_n , \quad 
P = \sum_{n=2}^\infty n R_n . 
\lbl{PQdef} \eeq 
Applying the law of mass action to (\ref{Sbreakdown})--(\ref{poisoning}) 
and using the definitions of (\ref{PQdef}), we  obtain the kinetic equations 
\beqa
\frac{d S}{dt} & = & S_0 - 2\ep S - k S(P+Q) \lbl{5Sdot} \\ 
\frac{d L_1}{dt} &=& \ep S + \half k S [(1+f)Q+(1-f)P] - 
				a L_1 (2L_1+L+\chi R) \\ 
\frac{d L_n}{dt} &= & a L_1 ( L_{n-1} - L_n ) - a\chi L_n R_1 
				\qquad (n\geq2)\lbl{Lndot}\\ 
\frac{d R_1}{dt} &= & \ep S + \half k S [(1+f)P+(1-f)Q] - 
				a R_1 (2R_1+R+\chi L) \\ 
\frac{d R_n}{dt} &= & a R_1 ( R_{n-1} - R_n ) - a \chi R_n L_1 
				\qquad (n\geq2).\lbl{Rndot}
\eeqa
This constitutes a quantitative description of the polymerising system. 
Equations (\ref{Lndot}) and (\ref{Rndot}) hold for all integers $n\geq2$, 
that is for all polymer lengths, from dimers ($n=2$) to infinitely long polymers. 
A diagrammatic summary of the rate processes involved is shown in 
Figure \ref{detail-fig}. 

A considerable advantage of the present model is that it is 
possible to reduce the complexity of the corresponding infinite 
set of rate equations (\ref{5Sdot})--(\ref{Rndot}), resulting in 
a closed system of only seven equations which 
contains the dynamics of the full system. This is based on 
the quantities $S,L_1,R_1,L,R,P,Q$ which evolve according to 
\beqa
\frac{d S}{dt} & = & S_0 - 2\ep S - k S(P+Q) \lbl{7odeS} \\ 
\frac{d L_1}{dt} & = & \ep S + \half k S [(1+f)Q+(1-f)P] 
	- a L_1 (2L_1+L+\chi R) \lbl{L1doteq} \\
\frac{d L}{dt} & = & a L_1^2 - a\chi L R_1 \lbl{7odeL} \\ 
\frac{d Q}{dt} & = &  2aL_1^2 +aLL_1-a\chi QR_1 \\ 
\frac{d R_1}{dt} & = & \ep S + \half k S [(1+f)P+(1-f)Q] 
	- a R_1 (2R_1+R+\chi L) \lbl{R1doteq} \\
\frac{d R}{dt} & = & a R_1^2   - a\chi R L_1  \lbl{7odeR} \\
\frac{d P}{dt} & = & 2aR_1^2 + a RR_1 -a\chi PL_1 . \lbl{7odeP}
\eeqa

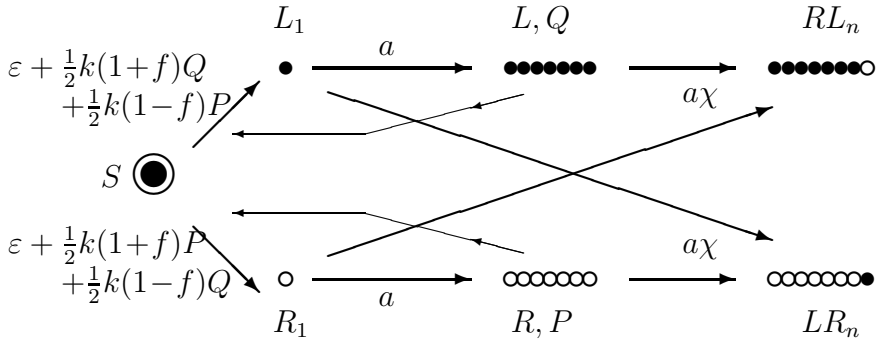
\begin{figure}[ht]
\begin{picture}(500,120)(-5,-12)
\thicklines
\put(50,50){\circle{15}}
\put(50,50){\circle*{10}}
\put(30,45){$S$}
\put(65,60){\vector(1,1){25}}
\put(65,30){\vector(1,-1){25}}
\put(-5,87){$\ep+\half k(1\!+\!f)Q$}
\put(15,72){$+\half k(1\!-\!f)P$}
\put(-5,20){$\ep+\half k(1\!+\!f)P$}
\put(15,5){$+\half k(1\!-\!f)Q$}
\put(100,90){\circle*{5}}
\put(100,10){\circle{5}}
\put(95,105){$L_1$}
\put(95,-10){$R_1$}
\put(110,90){\vector(1,0){60}}
\put(110,10){\vector(1,0){60}}
\put(135,95){$a$}
\put(135,0){$a$}
\multiput(185,90)(5,0){7}{\circle*{5}}
\multiput(185,10)(5,0){7}{\circle{5}}
\put(185,105){$L,Q$}
\put(185,-10){$R,P$}
\put(230,90){\vector(1,0){40}}
\put(230,10){\vector(1,0){40}}
\multiput(285,90)(5,0){7}{\circle*{5}}
\multiput(285,10)(5,0){7}{\circle{5}}
\put(320,90){\circle{5}}
\put(320,10){\circle*{5}}
\put(295,105){$RL_n$}
\put(295,-10){$LR_n$}
\put(116,19){\vector(3,1){168}}
\put(116,81){\vector(3,-1){168}}
\put(250,77){$a\chi$}
\put(250,20){$a\chi$}
\thinlines
\put(190,80){\line(-4,-1){60}}
\put(190,20){\line(-4,1){60}}
\put(190,80){\vector(-4,-1){20}}
\put(190,20){\vector(-4,1){20}}
\put(130,65){\vector(-1,0){50}}
\put(130,35){\vector(-1,0){50}}
\end{picture}
\caption{Illustration of our homochiral polymerisation scheme 
including the nonlinear feedback mechanisms. The quantities 
$LR_n$, $RL_n$ are not included in the mathematical model 
since they play no role in the reaction scheme. The quantities 
$L,R$ refer to the total numer of polymers of each handedness, 
and $P,Q$ to the total mass of material in polymeric form.  
It is these latter quantities ($P$, $Q$) which determine the 
effectiveness of the catalytic breakdown of the source species 
into monomers.}
\label{scheme-figfull}
\end{figure}

The rate processes described in terms of equations 
(\ref{7odeS})--(\ref{7odeP}) are depicted in Figure \ref{scheme-figfull}. 
This reduction from an infinite system of coupled  ordinary differential 
equations to a system of just seven ordinary differential equations is 
remarkable in that it is an {\em exact} simplification -- no approximations have 
been made. Such an exact reduction would not be possible if the 
growth rate coefficients were dependent on polymer length.   
If appropriate at all it would then have to rely on approximations, 
and an accurate approximation may require not just the zeroth and 
first moments of the distributions $L_n$, $R_n$, but on higher moments 
as well.  

%-----------------------
\subsection{Transforming the system of kinetic equations}

Before we analyse the steady-state kinetics, 
it is useful to recast the system of seven ordinary differential 
equations in an alternative form.  Instead of describing the 
concentration of each species separately, we assign variables 
for the total concentrations of source material ($S$),  monomers 
($\mu$), sequences ($N$) and sequence mass ($M$), 
\beq
\mu=L_1 + R_1 , \qquad N = L+R , \qquad M = P+Q , 
\lbl{muMNdef} \eeq
and a set of variables $\delta,\theta,\eta$ for the proportions of 
right-handed molecules in each of $\mu,N,M$ respectively  
\beq
\de = \frac{R_1-L_1}{R_1+L_1} , \qquad 
\theta = \frac{R-L}{R+L} , \qquad \eta = \frac{P-Q}{P+Q} .  
\lbl{dedef}\eeq
These quantities relate to the chiral purity of the system: 
$\de$ is the chiral purity of the monomers ($\mu$), 
$\theta$ and $\eta$ describe the chiral purity of the homochiral polymer chains, 
$\theta$ being a number-weighted measure (corresponding to $N$) and 
$\eta$ a mass-weighted measure (corresponding to $M$). 

We transform the kinetic equations (\ref{7odeS})--(\ref{7odeP}) 
into the variables relating to the concentrations of polymers given by 
(\ref{muMNdef})--(\ref{dedef}) and hence obtain the seven coupled equations 
\beqa
\frac{d S}{dt} & = & S_0 - 2\ep S - k S M , \lbl{kinS} \\ 
\frac{d \mu}{dt} & = & 2\ep S + k S M - a\mu^2 (1+\de^2) -
\half a\mu N(1+\de\theta)-\half a\chi\mu N(1-\de\theta),\nn\\&&\lbl{kinmu}\\ 
\frac{d N}{dt} & = & \half a\mu^2(1+\de^2) - 
\half a \chi \mu N (1-\de\theta) , \\ 
\frac{dM}{dt} & = & a\mu^2(1+\de^2) + \half a \mu N (1+\de\theta) - 
\half a \chi \mu M (1-\de\eta) , \lbl{kinM} \\ 
\frac{d\eta}{dt} & = & \frac{a\mu^2}{M}(2\de-\eta-\eta\de^2) + 
\frac{a\mu N}{2M} (\de+\theta-\eta-\de\theta\eta) + 
\half a \chi \mu \de (1-\eta^2) , \nn\\&&\lbl{kine} \\ 
\frac{d\theta}{dt} & = & \frac{ a \mu^2}{2N}(2\de-\theta-\theta\de^2) + 
\half a \chi \mu \de (1-\theta^2) , \lbl{kinth} \\ 
\frac{d\de}{dt} & = & -\frac{2\ep S\de}{\mu} - \frac{kSM\de}{\mu} + 
\frac{kfSM\eta}{\mu} - \half a (1-\de^2)(2\mu\de+N\theta-\chi N\theta) . 
\nn\\ && \lbl{kinde} \eeqa
The advantage of (\ref{kinS})--(\ref{kinde}) over 
(\ref{7odeS})--(\ref{7odeP}) lies in the considerable reduction in the amount of 
algebra required to derive solutions; for example it is easy to see that 
$\de=\theta=\eta=0$ satisfies (\ref{kine})--(\ref{kinde}), leaving 
a system of four equations (\ref{kinS})--(\ref{kinM}) for the four 
unknowns $S,\mu,N,M$.   

The equations (\ref{kinS})--(\ref{kinde}) 
will be used throughout the rest of the paper.  In later sections 
we shall consider the equations for the symmetric growth of 
homochiral polymer sequences, (\ref{kinS})--(\ref{kinM}), 
separately from the equations describing the chiral purity 
of the system,  (\ref{kine})--(\ref{kinde}).

%----------------------
\section{Steady-state behaviour}
\label{ss-sec}

For the analysis of the steady-state solutions all the right-hand sides 
of equations (\ref{7odeS})--(\ref{7odeP}), or equations 
(\ref{kinS})--(\ref{kinde}) are set to zero. 
For the later stages of the calculation of the steady-states 
we ignore the small parameter $\ep$ (that is, we set $\ep=0$), 
since this will allow explicit analytical formulae to be derived 
and this will not greatly influence the steady-states.  
The reason for this is that once there are appreciable numbers of 
polymers of either handedness present in the system, the catalytic 
breakdown of source will dominate the production of monomers of 
both handednesses. The ${\cal O}(\ep)$ terms in equations 
(\ref{7odeS})--(\ref{7odeP}) will be reinstated later where the 
kinetics of  the system starting from zero initial data are investigated. 

We solve the equations (\ref{7odeS}), (\ref{7odeL}) and 
(\ref{7odeR}) to express the solution in terms of $L_1,R_1$
\beq
L = \frac{L_1^2}{\chi R_1} , \quad 
R = \frac{R_1^2}{\chi L_1} , \quad 
P = \frac{R_1^2(R_1+2\chi L_1)}{\chi^2 L_1^2} , \quad
Q = \frac{L_1^2(L_1+2\chi R_1)}{\chi^2 R_1^2} , 
\eeq
\beq
S = \frac{S_0 \chi^2 L_1^2 R_1^2}{2\ep \chi^2 L_1^2 R_1^2  + 
k (L_1^5 + R_1^5 + 2\chi L_1^4R_1 + 2\chi L_1 R_1^4)} . 
\lbl{LRSPQ-sss} \eeq 
The microscopic steady-state solution is then found to be 
\beq
L_n = L_1 \left( \frac{L_1}{L_1+\chi R_1} \right)^{n-1} , \qquad 
R_n = R_1 \left( \frac{R_1}{R_1+\chi L_1} \right)^{n-1} . 
\eeq
In terms of the variables $N$, $M$, $\theta$, $\eta$, 
the steady-state solution is then 
\beq
N = \frac{\mu(1+3\delta^2)}{\chi(1-\de^2)} , \quad 
M = \frac{\mu [ 1 \!+\!10 \de^2 \!+\! 5\de^4 + 
2\chi(1\!-\!\de^2)(1\!+\!3\de^2) ] }{\chi^2(1-\de^2)^2} , 
\eeq
together with 
\beq
S = \frac{S_0 \chi^2 (1-\de^2)^2}{ 2\ep\chi^2(1-\de^2)^2 + 
k\mu[1+10\de^2+5\de^4+2\chi(1\!-\!\de^2)(1\!+\!3\de^2)]} . 
\eeq
The average chain length is given by $M/N$, which can 
be written as 
\beq
\frac{M}{N}=2+\frac{1+10\de^2+5\de^4}{\chi(1-\de^2)(1+3\de^2)} . 
\eeq
The most important thing to notice about the second term here is that 
asymmetric (chiral, i.e. $\de\neq0$) solutions permit longer chains to 
be produced. For the symmetric solution ($\de=0$) we have $M/N=2+1/\chi$; 
whereas if $\de$ approaches $\pm1$ then arbitrarily large chains 
can be produced.  We note that large inhibition rates ($\chi$) reduce the 
expected sequence length (as one might expect); however, 
we shall show later on that this effect is compensated for since larger 
values for $\chi$ make it easier for the system to adopt a chiral state.

\begin{figure}[t]
\vspace{60mm}
\includegraphics{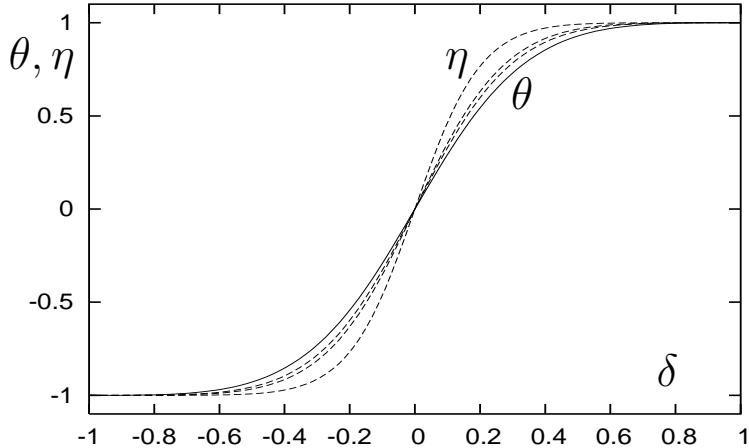}
\begin{picture}(0,0)(0,0)
\put(260,25){{\Large $\delta$}} 
\put(15,145){{\Large $\theta,\eta$}}
\put(205,130){{\Large $\theta$}}
\put(180,145){{\Large $\eta$}}
\end{picture}
\caption{Graph of steady-state values for the chiral purities 
($\theta,\eta$) of the homochiral sequences against the chiral 
purity of the monomer, $\de$. These are given by equation 
(\protect\ref{th-et-de-sss}), and show the accentuated effect 
of symmetry-breaking in the homochiral sequences compared 
to the monomers. The solid line corresponds to $\theta$; 
the dashed lines refer to $\eta$.  The steepest curve relates to $\chi=0$, 
the next to $\chi=1$ and that for $\chi=2$ is the closest 
to the $\theta$ curve. }
\label{th-et-fig}
\end{figure}

It is more natural to express solutions in terms of $\mu,\delta$ 
than $L_1,R_1$.  The chiral purity of the polymers can be 
defined in two ways, one by the mass-weighted purity ($\eta$) 
and the other in terms of the number-weighted purity ($\theta$).  
At steady-state we have 
\beq
\theta = \de \left( \frac{3+\de^2}{1+3\de^2} \right) , \qquad 
\eta = \de \left( \frac{5+10\de^2+\de^4 + 2 \chi (1-\de^2)(3+\de^2)}
{1+10\de^2+5\de^4 + 2 \chi (1-\de^2)(1+3\de^2)} \right) . 
\lbl{th-et-de-sss} \eeq 
Thus if the chiral purity of the monomer, $\de$, departs from zero, 
then the chiral purity of the homochiral sequences ($\theta$ and $\eta$) 
also do, and by greater amounts, as illustrated in Figure \ref{th-et-fig}.  
This can be seen from the linearisations for small $\de$ which 
give $\theta=3\de$ and $\eta=(5+2\chi)\de/(1+2\chi)$. 
For large $\chi$ the curve for $\eta$ approaches that for $\theta$. 

When substituted into equation (\ref{kinmu}) the above equations yield 
\beq
\mu^2 = \frac{2\chi S_0 (1-\de^2)}{3a\chi (1-\de^4) + 
a (1+6\de^2+\de^4)} . \lbl{mueq}
\eeq
Finally we have to determine the chiral purity of the 
monomers, $\delta$. This is the most complicated part of the 
calculation so, for the sake of clarity, we now set $\ep=0$; 
$\delta$ is then given by either $\delta=0$, the symmetric 
(achiral) steady-state solution or, in terms of the fidelity, by 
\beq
f = \left( \frac
{1\!+\!10\de^2\!+\!5\de^4 + 2\chi(1\!-\!\de^2)(1\!+\!3\de^2)}
{5\!+\!10\de^2\!+\!\de^4+2\chi(1\!-\!\de^2)(3\!+\!\de^2)} 
\right) \left( \frac{4(1\!+\!\de^2) + 2\chi(1\!-\!\de^2)}
{1\!+\!6\de^2\!+\!\de^4 + 3\chi (1\!-\!\de^4)} \right) . 
\lbl{fde-def} \eeq
We shall discuss this chiral solution in further detail later 
on (Section \ref{asym-sss-sec}). 
One question we aim to address in the remainder of this section is 
how small $f$ could be, and the system still exhibit a bifurcation 
to a non-symmetric state.  

\begin{figure}[t]
\vspace{60mm}
\includegraphics{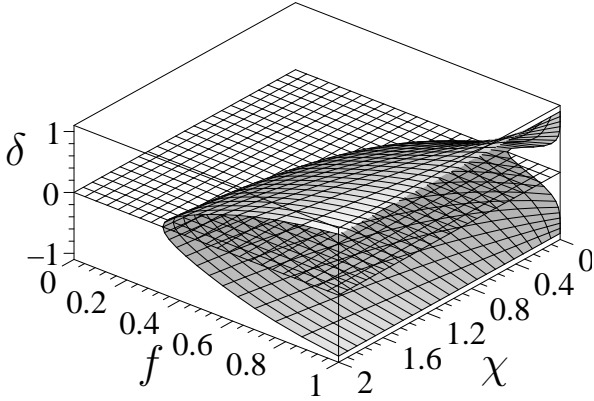}
\begin{picture}(0,0)(0,0) 
\put( 40,90){{\Large $\delta$}}
\put( 90,10){{\Large $f$}}
\put(220,10){{\Large $\chi$}}
\end{picture} 
\caption{Graph of the chiral purity $\de$ against $\chi$ and $f$ 
for the steady-state values given by equation (\protect\ref{fde-def}), 
for fidelity in the range $0<f<1$ and with $0<\chi<2$. }
\label{bif-fig}
\end{figure}

%--------------
\subsection{The symmetric, achiral, steady-state solution}
\label{sec31}

It is not immediately obvious that all the solutions given above 
in equations (\ref{LRSPQ-sss})--(\ref{fde-def}) for the steady-state 
exist for all parameter values, or whether any 
particular solution is unique.  Hence we start by considering 
the symmetric solution, where $\de=\theta=\eta=0$. 
This solution exists for all parameter values and in this case we have 
\beq 
\mu = \sqrt{\frac{2\chi S_0}{a(1+3\chi)} } , \qquad 
S = \frac{\chi\sqrt{a\chi S_0(1+3\chi)}}{\sqrt{2} k(1+2\chi)} , 
\lbl{ssss-mu-S} \eeq 
\beq 
N=\frac{\mu}{\chi}= \sqrt{\frac{2 S_0}{a\chi(1\!+\!3\chi)}},\quad 
M = \frac{(1\!+\!2\chi)\mu}{\chi^2} = 
\frac{(1\!+\!2\chi)\sqrt{2S_0}}{\chi  \sqrt{a\chi(1\!+\!3\chi)}} , 
\lbl{ssss-NM} \eeq 
with $L_1=R_1=\half \mu$, $L=R=\half N$, $P=Q=\half M$. 

We now consider the mathematical stability of this solution, 
that is we aim to answer the question: `If the system 
is close to the steady-state solution, will it be attracted closer to it, 
or diverge further away from it ?'   To answer this question, we linearise 
around the solution $\de=\theta=\eta=0$.  Note that the 
formulae for $\mu,N,M,S$ all have ${\cal O}(\de^2)$ correction 
terms, and no ${\cal O}(\de)$ terms.  Thus these will 
be treated as constants. 
We only need to analyse the evolution of $(\theta,\delta,\eta)$ 
over time, starting from small perturbations away from (0,0,0).
Linearising equations (\ref{kine})--(\ref{kinde}) we obtain 
\beqa
\frac{d \eta}{dt} & = & \frac{a\mu^2}{M}(2\de-\eta) + 
\frac{a\mu N}{2M} (\de+\theta-\eta) + \half a \chi \mu \de
\lbl{lin-eta} \\ 
\frac{d \theta}{dt} & = & \frac{ a \mu^2}{2N}(2\de-\theta) + 
\half a \chi \mu \de \\ 
\frac{d \de}{dt} & = & %-\frac{2\ep S\de}{\mu} 
- \frac{kSM\de}{\mu} + \frac{kfSM\eta}{\mu} - 
\half a (2\mu\de+N\theta-\chi N\theta) . \lbl{lin-de}
\eeqa
Note that we have ignored the ${\cal O}(\ep)$ term.
Inserting the expressions for $\mu$, $N$ and $M$ from 
(\ref{ssss-mu-S})--(\ref{ssss-NM}) we obtain 
\beq
\frac{d}{dt} \left( \begin{array}{c} \eta \\ \theta \\ \delta 
\end{array} \right) = 
\half a\mu \chi \left( \begin{array}{ccc} -1 & \frac{1}{(1+2\chi)} &
\frac{2(1+3\chi)}{(1+2\chi)} \\ 0 & -1 & 3 \\ 
\frac{f(1+3\chi)}{\chi^2} &
\frac{\chi-1}{\chi^2} & -\frac{(1+5\chi)}{\chi^2} 
\end{array} \right)  
\left( \begin{array}{c} \eta \\ \theta \\ \delta \end{array} \right) ,
\lbl{stab-mat} \eeq
which has the characteristic polynomial 
\beqa
0 & = & \chi^2\lambda^3 + (2\chi^2+5\chi+1)\lambda^2 + 
\left(\chi^2+7\chi+5 - \frac{2f(1+3\chi)^2}{1+2\chi}\right)\lambda + 
\nn\\ && + \left(4+2\chi - \frac{f(5+6\chi)(1+3\chi)}{(1+2\chi)}\right) . 
\lbl{lam-poly} \eeqa
The Routh-Hourwitz criteria state that all solutions of 
$\lambda^3+A\lambda^2+B\lambda+C=0$ satisfy $\Re(\lambda)<0$ 
if and only if $A>0$, $C>0$ and $AB>C$ (for further details, see Murray, 
1989). If all values of $\Re\lambda$ are negative then any 
solution of (\ref{stab-mat}) will have all the quantities $\eta$, $\theta$ 
and $\delta$ decaying to zero as time increases.  In our example, 
clearly $A>0$ whatever values $\chi$ and $f$ 
take, but the other two conditions are less clear. The condition $C>0$ 
gives $f>f_c(\chi)$ with $f_c(\chi)$ given by 
\beq
f_c(\chi) = \frac{2(2+\chi)(1+2\chi)}{(5+6\chi)(1+3\chi)} . 
\lbl{fcdef} \eeq
This value for $f_c$ agrees with $f$ as given by (\ref{fde-def}) in the 
case $\delta=0$. The condition $AB>C$ implies an instability when 
$f>f_{c2}(\chi)$; however, $f_{c2}(\chi)$ lies in the region $f>1$ 
for all $\chi$ so this instability can be ignored, since only $f\leq1$ is 
physically realisable.   So the symmetric solution is stable for 
$f<f_c$, and unstable for $f>f_c$.   The value of $f_c$ lies between 
$2/9$ and $4/5$ and depends on $\chi$: at $\chi=0$; the achiral solution 
is unstable for  $f>4/5$, whereas at large $\chi$, 
the instability of the achiral solution occurs for $f>2/9$. 

%--------------
\subsection{The chiral (asymmetric) steady-state solution}
\label{asym-sss-sec}

Once $f>f_c(\chi)$ we have the existence of two chiral steady-states 
as well as the achiral steady-state. Furthermore, since the Routh-Hourwitz 
criteria are `necessary and sufficient' for the existence of a solution of
(\ref{lam-poly}) with $\Re(\lambda)<0$, once $f>f_c$ the achiral solution 
becomes unstable, so will not be observed in any physical system. 
Thus once the chiral solutions exist, they become the `preferred' 
steady-state solutions since they are mathematical stable. 
Even if a system were artificially put into the symmetric (achiral) state, 
any small perturbation would cause some chiral imbalance and the natural 
kinetics of the model would then carry the whole system to one 
of the two steady chiral states in which one handedness 
of homochiral polymers dominates the other. 

At $f=f_c$ a bifurcation occurs, and two mirror-image chiral solutions appear.  
We now briefly examine the neighbourhood of this point in more detail. 
These chiral steady-states are governed by equation (\ref{fde-def}); from 
this equation we note $\de=0$ 
implies $f=f_c$.   Assuming small $\de$ and expanding we find 
$f=f_c + \beta\de^2$ with $\beta>0$ so the bifurcation is supercritical, 
and the new solutions exist in $f>f_c$ where the achiral solution is unstable. 
Thus, as expected, the bifurcation occurs at exactly the same position as 
that at which the achiral steady-state becomes unstable.  
This is a standard result in bifurcation theory; see Guckenheimer \& 
Holmes (1983) or Berg\'{e} \etal\ (1984) for more details. 
Although this is a curve in $(f,\chi)$ space, we shall consider $f$ to be the 
primary bifurcation parameter, with the bifurcation point $f_c$ depending on $\chi$. 

\begin{figure}[t]
\vspace{53mm}
\includegraphics{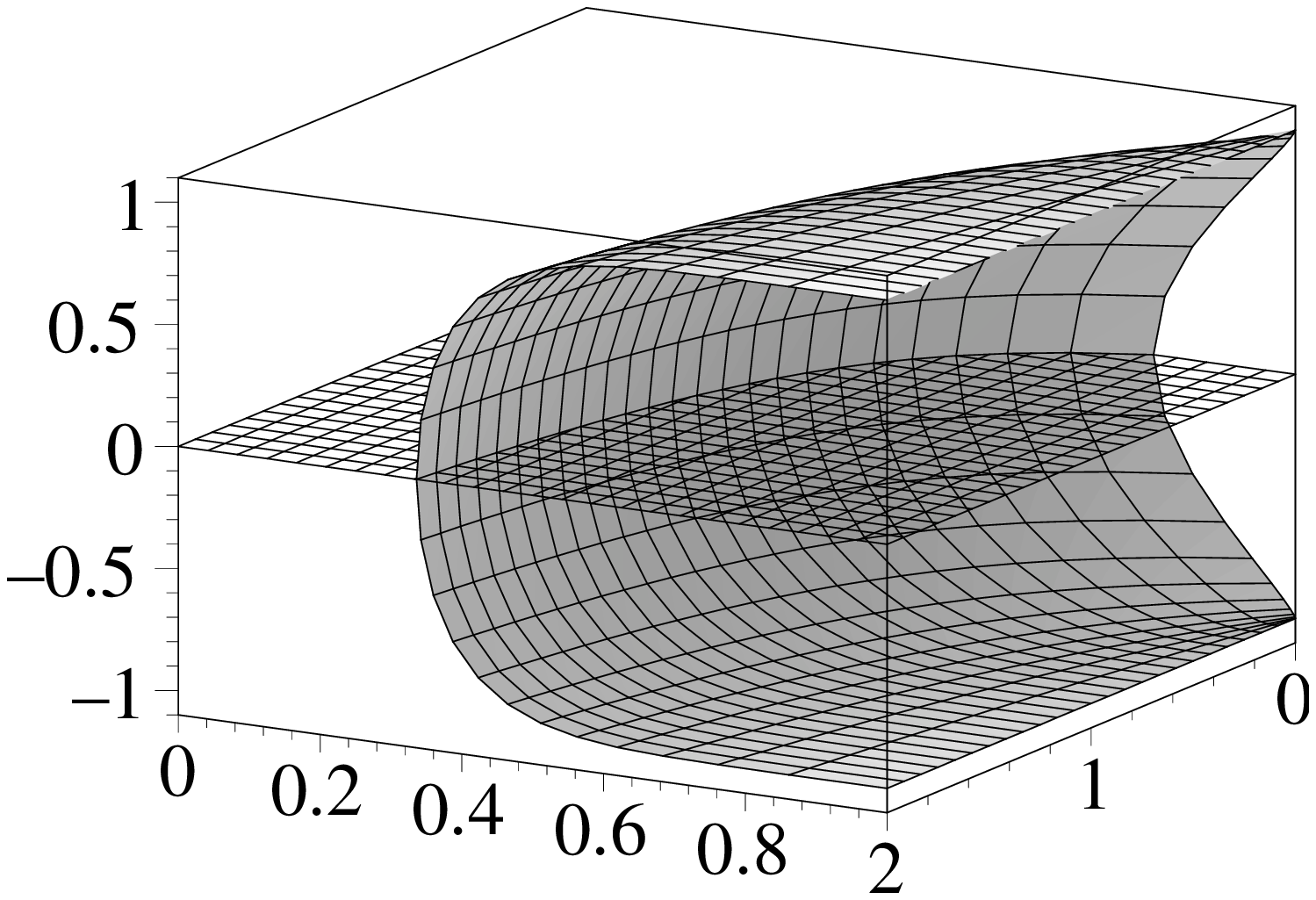}
\includegraphics{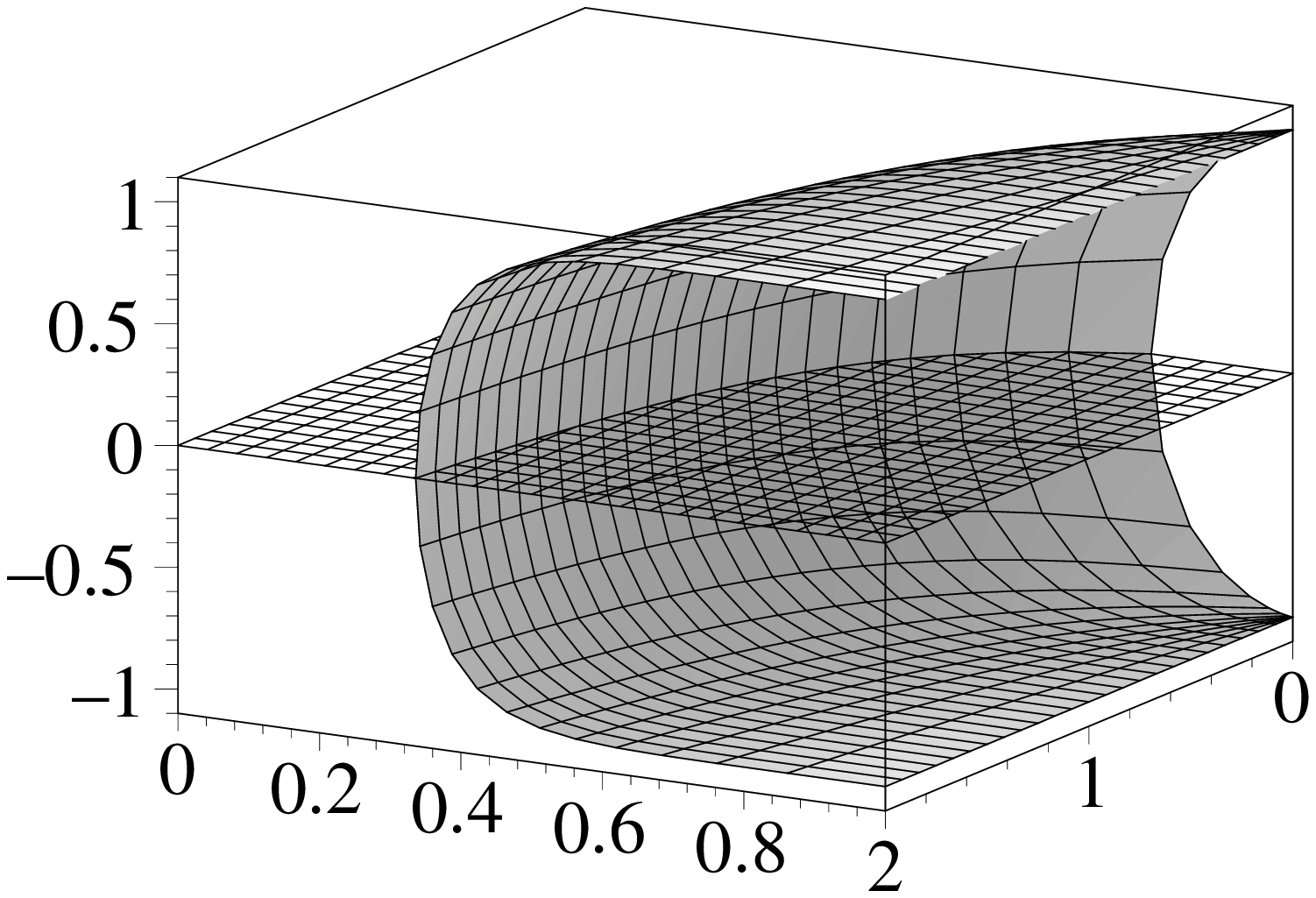}
\begin{picture}(0,0)(0,-5) 
\put( -5,99){{\Large $\theta$}} 
\put( 31, 0){{\large $f$}} % Correct coords on f,\chi, edit both .ps files
\put(130,10){{\Large $\chi$}}
\put(170,99){{\Large $\eta$}} 
\put(209, 0){{\large $f$}}
\put(305,10){{\Large $\chi$}} 
\end{picture} 
\caption{Graph of the chiral purities $\theta$ (on the left) 
and $\eta$ (right) against $\chi$ and $f$ for the steady-state 
values given by equation (\protect\ref{fde-def}).  This is to be compared 
with Figure \protect\ref{bif-fig}, illustrating that the 
polymer concentration and mass show more extreme 
chiral purities than the monomers (described by $\de$).}
\label{bif-fig2}
\end{figure}

In figure \ref{bif-fig} we show the chiral steady-state 
solutions given by (\ref{fde-def}) parametrically, plotting 
$\delta$ against $f$ and $\chi$. The bifurcation occurs at 
$f_c=4/5$ if $\chi=0$, reducing to $f_c = 2/9$ in the limit of 
large $\chi$. Thus we see the beneficial effect of sequence inhibition 
if one wants a system which undergoes a symmetry-breaking
bifurcation at small values of the fidelity parameter $f$. 
Figure \ref{bif-fig2} shows the chiral purities of the 
polymers, firstly weighted by number ($\theta$) and then 
by mass ($\eta$).  Both show that the chiral purity of the 
polymers is much greater than that of the monomer 
(compare Figure \ref{bif-fig2} with Figure \ref{bif-fig}). 
Note also that for the range of fidelities $2/9<f<4/5$ 
the bifurcation to asymmetric (chiral) solutions can 
occur by increasing $\chi$. 

Another natural question to pose here is whether only small and moderate values 
of $\de$ can be accessed; we put $\nu=1-\delta$ with $\nu\ll1$ to 
determine under what conditions $\de$ can approach unity. 
We find from equation (\ref{fde-def}) that $f=1-\chi\nu$. 
In this case we also expect $\theta$ and $\eta$ to be near unity. 
In fact the chiral purity of the polymeric sequences is much enhanced 
over the chiral purity of monomers, since a two-term expansion of 
equation (\ref{th-et-de-sss}) gives 
\beq
\theta \sim 1 - \rec{4} \nu^3 , \qquad 
\eta \sim 1 - \rec{4} \chi \nu^4 ,  
\eeq
for $\nu\ll1$ (in the special case $\chi=0$ we have 
$\eta\sim1-\rec{16}\nu^5$). 
The concentrations scale with $\nu$ according to 
\beq
R_1 =\sqrt{\frac{S_0 \chi \nu}{2a}} , \qquad 
R = \sqrt{\frac{2S_0}{a\chi\nu}} , \qquad 
P = \sqrt{\frac{8S_0}{a\chi^3\nu^3}} , \lbl{Rnuasy}
\eeq
\beq
L_1 = \sqrt{\frac{S_0\chi\nu^3}{8a}} , \qquad
L = \sqrt{\frac{S_0\nu^5}{32 a \chi}} , \qquad 
Q = \sqrt{\frac{2S_0 \nu^5}{a\chi}} , \lbl{Lnuasy}
\eeq
together with $S={\cal O}(\nu^{3/2})$.  

In summary we find that chiral solutions do indeed exist 
provided $f>f_c$ and this is easier to satisfy at larger inhibition 
rates ($\chi$).  Thus the presence of stronger cross inhibition aids 
the manifestation of chiral steady-states. 
Also, the higher the fidelity ($f$), the greater the dominance of one 
chirality over the other. 

We now turn to an analysis of the kinetics of sequence growth, 
in order to determine at what stage of the reaction the system 
is likely to manifest a chiral state. 

%----------------------------------------------
\section{Kinetics and stability of achiral solutions}
\label{kin-sec}

In this section we reintroduce the ${\cal O}(\ep)$ term omitted 
in the analysis following (\ref{th-et-de-sss}) and solve the 
kinetic problem for the achiral solution, in the limit $\ep\ll1$.   
We assume that initially there are no monomers ($\mu=0$), 
no polymers ($M=N=0$) and no precursor ($S=0$), 
though this source material is added continuously to the system 
starting at time $t=0$.  If $\ep=0$ then since there are no polymers, 
the precursor cannot be broken down and no polymers will ever form, 
so we need the ${\cal O}(\ep)$ term to produce some monomer. 
In section \ref{stab-sec} we analyse the stability 
of the solution, that is, whether small, random perturbations to 
such a solution grow or are damped out, as the total number 
of polymers and monomers grow from zero concentrations. 

%----------------------------
\subsection{Growth of the achiral solution}
\label{grow-sec}

We assume a set of initial conditions in which $\eta = \theta = 0 = \de$, 
and then $\eta = \theta=0=\de$ for all subsequent times.  This reduces 
the system (\ref{kinS})--(\ref{kinde}) to a system of three ordinary 
differential equations which we solve asymptotically in the limit $\ep\ll1$. 
Our aim is to develop matched asymptotic expansions for the solution of 
\beqa
\frac{d S}{dt} & = & S_0 - 2\ep S - k S M , \\ 
\frac{d\mu}{dt} & = & 2\ep S + k S M - a \mu^2 - \half a \mu N (1+\chi) ,\\ 
\frac{d N}{dt} & = & \half a \mu^2  - \half a \mu \chi N ,  \\ 
\frac{d M}{dt} & = & a\mu^2 + \half a N\mu -\half a \chi M\mu , 
\eeqa
through a series of timescales. 

%-------------
\subsubsection{Timescale I: $t={\cal O}(\ep^{-1/5})$}

For this timescale the appropriate scalings are 
$t = \ep^{-1/5} t_1$ together with 
\beq
S = \ep^{-1/5} S_1 , \qquad \mu = \ep ^{3/5} \mu_1 , 
\qquad N = \ep N_1 , \qquad  M = \ep M_1 , 
\lbl{T1-scal} \eeq
thus this is a long induction time over which the leading order equations are 
\beq
S_1' = S_0 , \quad \mu_1' = 2 S_1 + k S_1 M_1 , \quad 
N_1' = \half a \mu_1^2, \quad M_1' = a \mu_1^2 , 
\eeq
where prime denotes $d/dt_1$. This system has the solution
\beq
S_1 = S_0 t_1 , \qquad N_1 = \half M_1 , \qquad 
M_1 = \frac{\mu_1'}{kS_0t_1} -\frac{2}{k} . 
\eeq
where $\mu_1$ is given by the solution of 
\beq
\mu_1'' - \frac{\mu_1'}{t_1} - S_0 k a t_1 \mu_1^2 = 0 , 
\eeq
which unfortunately is not explicitly available. 
However, we can see that the timescale ends abruptly  
with $\mu_1,N_1,M_1$ all diverging as 
$t_1\rightarrow t_{1c}$, according to 
\beq
\mu_1 \sim \frac{6}{ak S_0 t_{1c}(t_{1c}-t_1)^2} , \qquad 
M_1 \sim \frac{6}{ak^2S_0^2t_{1c}^2(t_{1c}-t_1)^3} , 
\eeq
for some constant $t_{1c}$.  These relationships help us 
determine the scalings relevant in the next timescale. 
In this timescale we have seen the accumulation of source material, 
but this is only slowly converted into monomers and chains, 
so both of these grow very slowly, causing a big build up of 
source material until, at the end of this timescale, we see 
the concentration of chains increase to the level where the 
catalytic mechanism becomes active and accelerates the 
formation of monomers and chains. 

%------------
\subsubsection{Timescale II: $t=\ep^{-1/5}t_{1c}+{\cal O}(\ep^{1/5})$}
\label{T2-sec}

In this timescale all quantities are large and evolve quickly. 
To be specific we have 
\beq
S = \ep^{-1/5} S_2 , \quad 
\mu = \ep^{-1/5} \mu_2 , \quad 
N = \ep^{-1/5} N_2 , \quad
M = \ep^{-1/5} M_2 , 
\eeq
together with $t = \ep^{-1/5} t_{1c} + \ep^{1/5} t_2$. Using 
primes to denote $d/dt_2$, the leading order equations are 
\beqa &&\begin{array}{lcl}
S_2' = - k S_2 M_2 , \quad &&
\mu_2 ' = k S_2 M_2 - a\mu_2^2- \half a (1+\chi) \mu_2 N_2 , 
\\[2ex]
N_2' = \half a \mu_2^2 - \half a \chi \mu_2 N_2 , \quad &&
M_2' = a\mu_2^2 + \half a \mu_2 N_2 - \half a\chi \mu_2 M_2 , 
\end{array} \nn\\ && \lbl{T2eqs} \eeqa
As the chains are present in large enough quantities for the catalytic 
mechanism to be active, and since there is a large amount of source 
material present at the start of this timescale,  
this source material is rapidly converted into monomers and 
chains so that $\mu_2$, $N_2$ and $M_2$ all increase 
at the expense of the source species, $S$, 
whose concentration now monotonically reduces, so that the only 
significant simplification in the equations is in the equation for $S_2$. 

No explicit solution which matches back into Timescale I is available; 
however, the form of the large-time solution in Timescale II can be 
determined. Consider new timescale given by 
$\frac{1}{\mu_2}\frac{d}{dt_2}=\frac{d}{d\tau}$, 
hence $t_2=\frac{2}{a} \int \frac{d\tau}{\mu_2(\tau)}$, then 
the system (\ref{T2eqs}) can be written as 
\beqa 
\frac{d\mu_2}{d\tau} & = & Q - 2 \mu_2 -(1+\chi) N_2 \lbl{t2-lin-mueq} \\
\frac{dN_2}{d\tau} & = & \mu_2 - \chi N_2 \lbl{t2-lin-Neq} \\
\frac{dM_2}{d\tau} & = & 2\mu_2 + N_2 - \chi M_2 , \lbl{t2-lin-Meq} 
\eeqa 
which is a linear system, together with the equation 
\beq
\frac{d}{d\tau} (\log S_2) = -\frac{2kM_2}{a\mu_2} , 
\lbl{logS2eq} \eeq
{}from which $Q=2k S_2 M_2 / a \mu_2$ is obtained. 
The solution of equations (\ref{t2-lin-mueq})--(\ref{t2-lin-Meq}) 
is given by a combination of a complimentary function which solves 
the system with $Q(t)=0$ and a particular solution which satisfies 
the $Q(t)$ input term in equation (\ref{t2-lin-mueq}). 
For general parameter values, the forms of these solutions 
cannot be explicitly determined, however we shall study the two 
particular limits of large and small $\chi$ in which approximations 
can be obtained.  These correspond respectively to the cases 
where the solution is dominated by the particular solution and the 
complimentary function at large times. 

%-------------------------------------------
\subsubsection{The solution in  timescale II for $\chi\gg1$}

If we assume that the input function $Q$ has the form 
$Q_0 \ee^{-\lambda\tau}$, then the particular solution has the form 
\beq 
\left(\!\begin{array}{c}\mu_2(\tau)\\N_2(\tau)\\M_2(\tau)\end{array}\right) =
\left(\!\begin{array}{c}\hat\mu_2\\ \hat N_2\\ \hat M_2 \end{array} \right) 
\ee^{-\lambda\tau} . 
\eeq 
In the calculation of $Q$, we then have $S_2\sim \ee^{-\lambda\tau}$ too, 
and this assumption implies $\lambda=2k\hat M_2 /a \hat\mu_2$, and 
$Q \sim \lambda \hat{S}_2 \ee^{-\lambda\tau}$.  In the calculation 
of the prefactors $\hat\mu_2,\hat N_2,\hat M_2$ we then have to solve 
a cubic, and there are constraints that all the prefactors must be positive. 

Inserting these into the equations (\ref{t2-lin-mueq})--(\ref{t2-lin-Meq}) 
we obtain 
\beq 
\hat N_2 = \frac{\hat\mu_2}{\chi-\lambda} , \qquad 
\hat M_2 = \frac{\hat\mu_2(2\chi-2\lambda+1)}{\chi-\lambda} , 
\eeq 
and the cubic equation for $\lambda$ is 
\beq 
\lambda^3 - 2\chi \lambda^2 + 
\left( \chi^2 + \frac{4k}{a} \right) \lambda - \frac{2k}{a}(1+2\chi)=0. 
\eeq 
For asymptotically large $\chi$, this has two roots near $\lambda=\chi$, 
and one near $\lambda=0$. 
The larger roots are at $\lambda=\chi \pm \sqrt{2k/a\chi}$, one of 
which violates the condition $\lambda<\chi$ and the other leads 
to a solution which rapidly decays in time.  The physically relevant 
solution corresponds to $\lambda \sim 4k/a\chi$, hence the solution 
\beq
( S_2 , \mu_2 , N_2 , M_2 ) \sim 
A \left( \frac{3a\chi^2}{4k} , \chi , 1 , 2 \right) \ee^{-4k\tau/a\chi} , 
\lbl{t2-bigchi-sol} \eeq
Since this result has been derived on the basis of large $\chi$, 
we see a slow exponential decay in $\tau$.  The complimentary 
function also decays exponentially in $\tau$, but with exponents 
$\lambda_1=3$, $\lambda_2=\chi-1$ and $\lambda_3=\chi$, 
hence the complimentary function decays much more rapidly than 
the particular solution. 

\begin{figure}[t]
\vspace{60mm}
\includegraphics{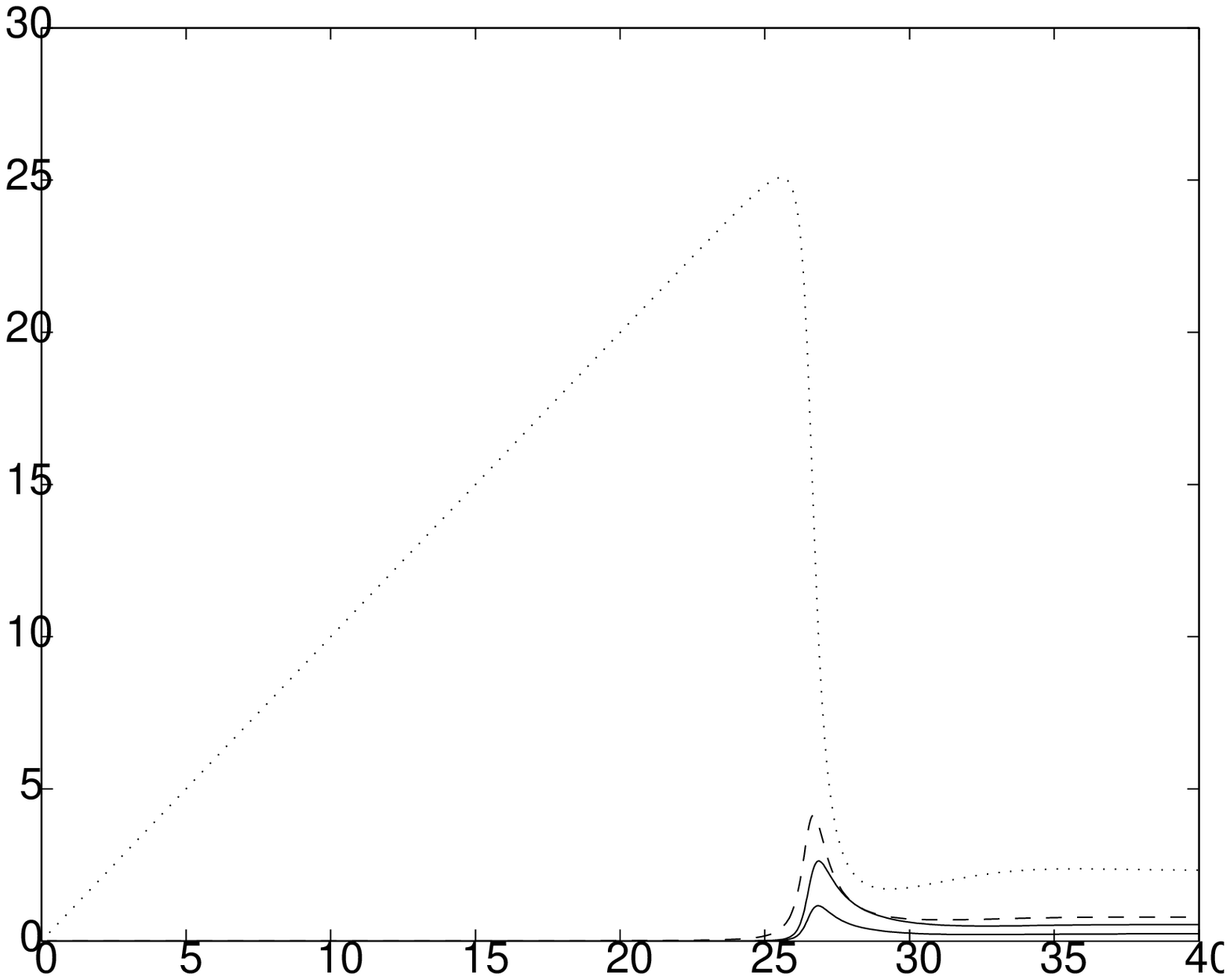}
\includegraphics{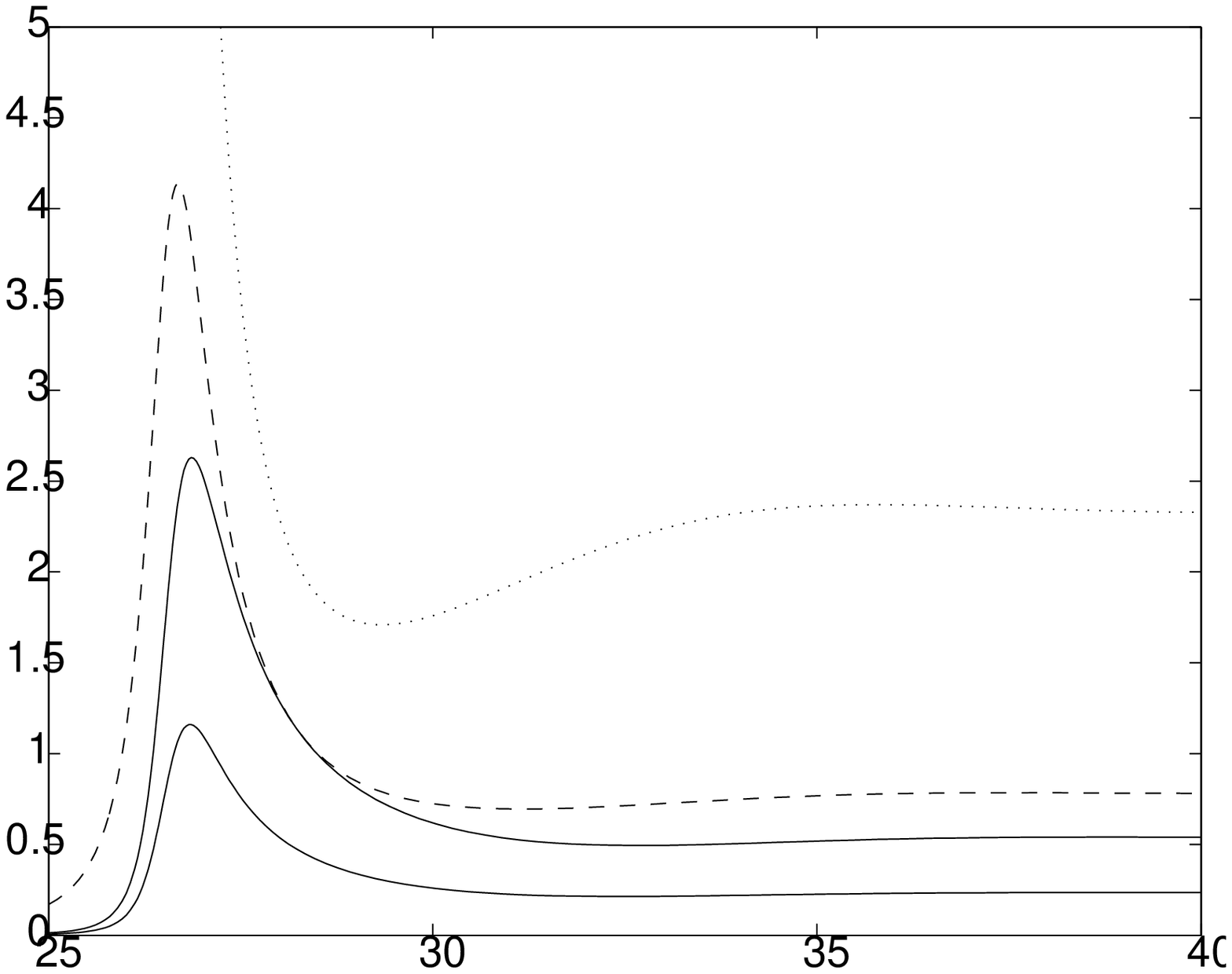}
\caption{Plot of a numerical solution of $S$, $\mu$, 
$N$ and $M$ against time for the case $\ep=10^{-5}$, 
$a=1$, $k=0.8$, $S_0=1$, $\chi=3.333$. The dotted 
curve corresponds to $S(t)$, the dsashed curve 
to $\mu(t)$ and the upper solid curve to $M(t)$ 
and the lower solid curve to $N(t)$. } 
\label{kin-big-fig}
\end{figure}

In terms of the original timescale $t_2$ the solution 
(\ref{t2-bigchi-sol}) leads to 
\beq
( S_2 , \mu_2 , N_2 , M_2 ) \sim \left( \frac{3a\chi^2}{8k^2t_2} , 
\frac{\chi}{2kt_2} ,  \frac{1}{2kt_2} , \frac{1}{kt_2} \right) . 
\eeq 
This agrees with the numerical observed results, which suggest that 
for $\chi\gg1 $ we observe that all quantities decay with $1/t_2$.  
Figure \ref{kin-big-fig} shows a numerical solution of the system 
produced by Matlab 6.5.0 Release 13; the right-hand graph shows in 
detail the decay of the concentrations at the end of the second timescale. 
When the concentrations $S,\mu,N,M$ become ${\cal O}(1)$ then other 
terms become relevant in the kinetic equations, and a further timescale is 
required to describe the final approach to equilibrium. 

%------------
\subsubsection{Timescale III for larger $\chi$: 
$t=\ep^{-1/5}t_{1c} + {\cal O}(1)$ }

As $S_2$, $\mu_2$, $N_2$ and $M_2$ all decay like 
$1/t_2$ at the end of the previous timescale, 
the new first term to become significant is the $S_0$ 
input into the $S$ equation. This becomes significant 
when $t_2=\ep^{-1/5}$, thus the third timescale 
is $t = \ep^{-1/5} t_{1c} + t_3$. In this timescale 
all of $S$, $\mu$, $M$ and $N$ are ${\cal O}(1)$. 
Thus the leading order equations are
\beqa
\frac{d S}{dt} & = & S_0 - k S M , \\ 
\frac{d \mu}{dt} & = & k S M - a \mu^2 - 
\half a \mu N (1+\chi) ,\\ 
\frac{d N}{dt} & = & \half a \mu^2  - \half a \mu \chi N ,  \\ 
\frac{d M}{dt} & = & a\mu^2 + \half a N\mu -\half a \chi M\mu , 
\eeqa
and over this timescale the system approaches its steady-state. 
Since at leading order the $\ep S$ terms are neglected, 
the leading order steady-state solution approached is 
precisely that described in Section \ref{ss-sec}. 

Overall, we see that there is a long induction time, of 
${\cal O}(\ep^{-1/5})$, followed by a some rapid kinetics during 
which the system explores states a long way from its steady-state, 
and then over a relatively fast (${\cal O}(1)$) timescale the system 
approaches its steady-state. We now derive the kinetics of the 
achiral solution for smaller $\chi$ before going to address the 
stability of the growing achiral solution. 

%------------
\subsubsection{The solution in timescale II for $\chi\ll1$}

\begin{figure}[t]
\vspace{60mm}
\includegraphics{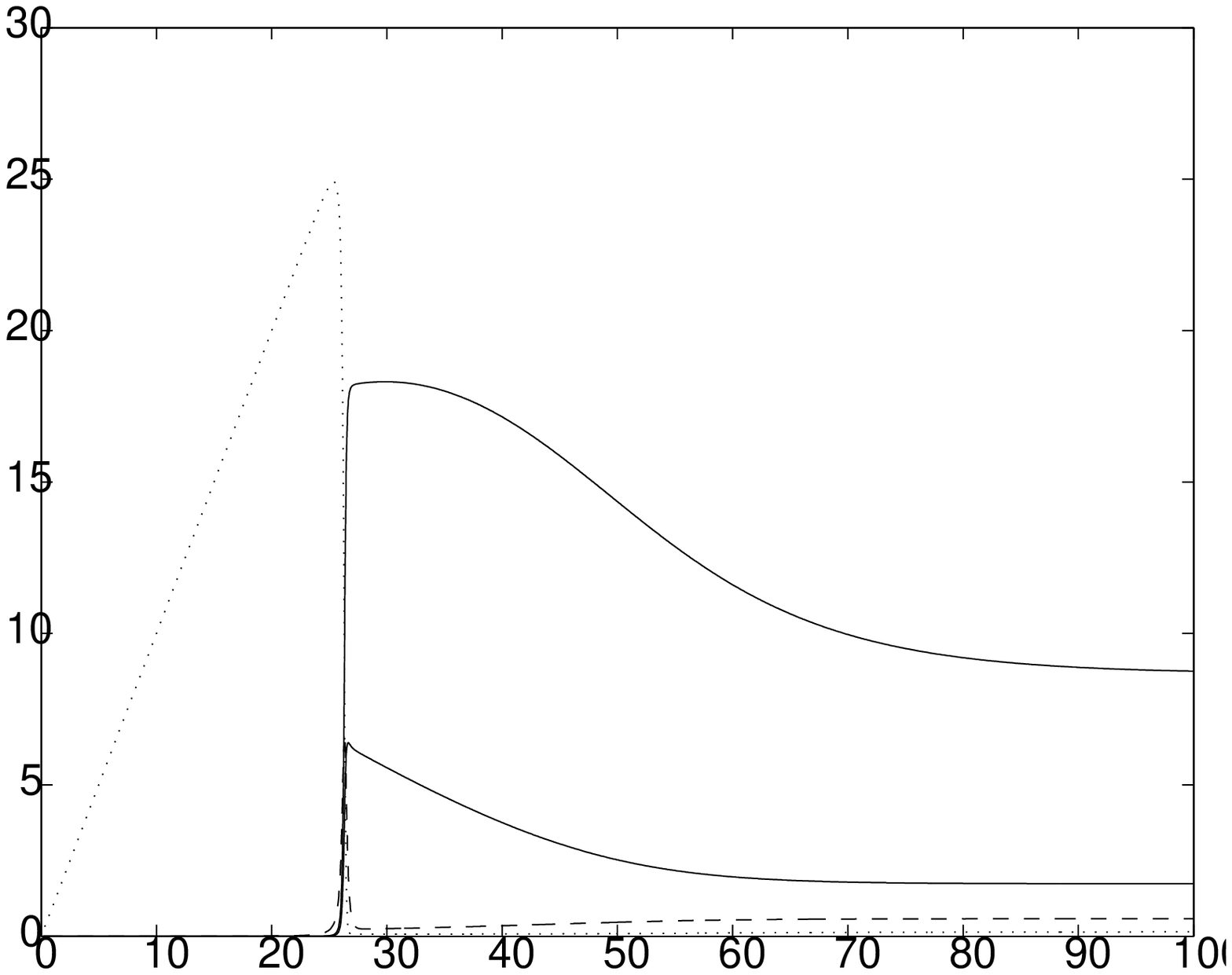}
\includegraphics{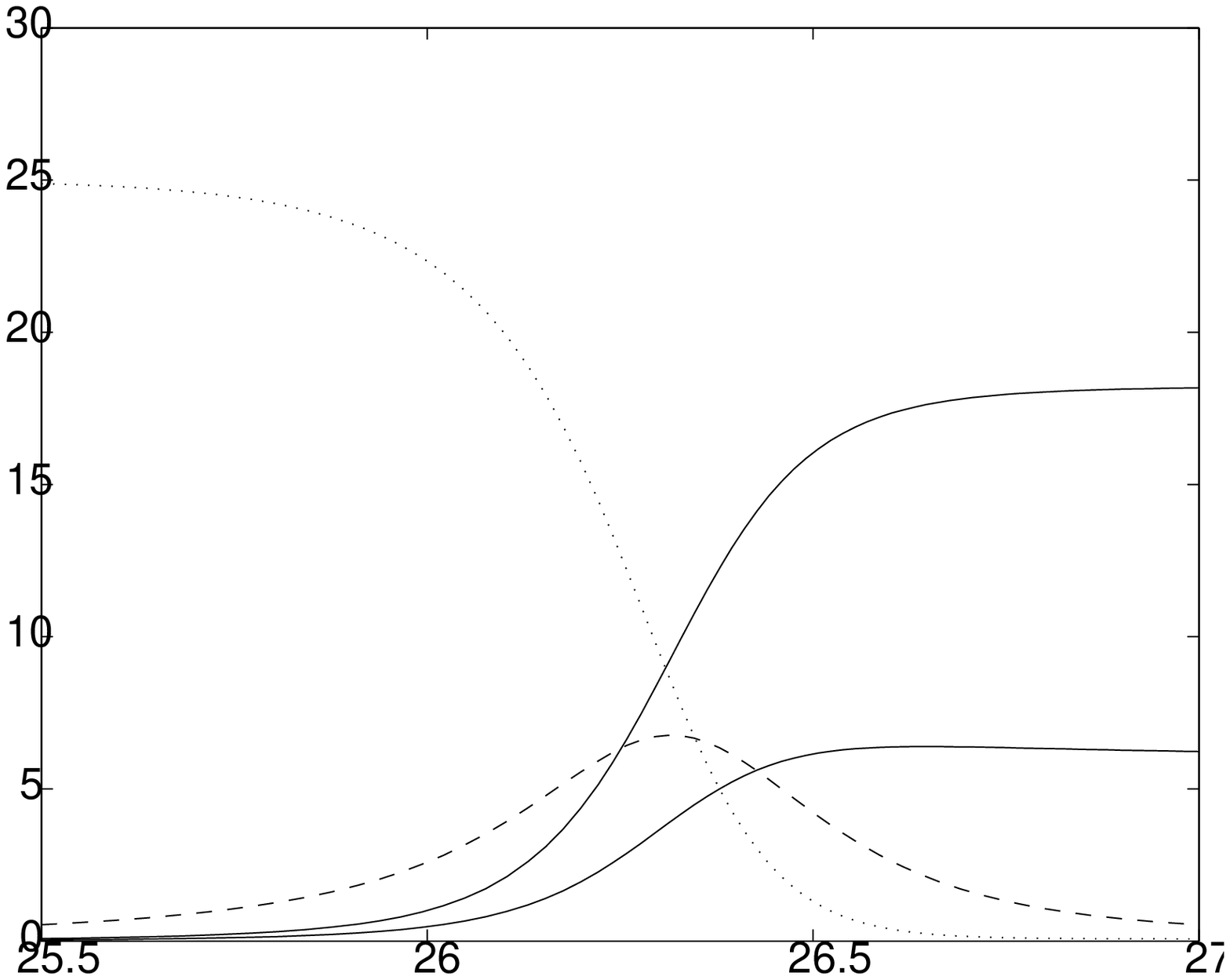}
\caption{Plot of numerical solution of $S(t)$ (the dotted curve), 
$\mu(t)$ (the dashed curve), $N(t)$ (the lower solid curve) 
and $M(t)$ (the upper solid curve). All parameters are as in 
Figure \protect\ref{kin-sm-fig} except $\chi=0.333$. }
\label{kin-sm-fig}
\end{figure}

Numerical simulations suggest that at the end of TII, $S_2$ and $\mu_2$ 
decay exponentially (in $t_2$) to zero, with $M_2$ and $N_2$ tending to 
constant values;  eventually reaching their ${\cal O}(1)$ steady-state values 
over subsequent and much longer timescales 
(see later subsections and Figure \ref{kin-sm-fig} for example). 

The linear system (\ref{t2-lin-mueq})--(\ref{t2-lin-Meq}) 
is solved by the sum of a particular solution and a complimentary function. 
For smaller $\chi$ the large-time solution is dominated by the complimentary 
solution component of the solution (i.e. that with $Q=0$) which is given by 
a combination of exponentials.  In the case $\chi\ll1$, the eigenvalues are 
$\lambda_1=-\chi$, $\lambda_{2,3} = -1 \pm i \sqrt{2\chi}$, thus applying 
the special result for repeated roots, at leading order, we have the 
complimentary function  
\beq
\left(\!\! \begin{array}{c} \mu_2\\N_2\\M_2 \end{array} \!\!\right) = 
C_1 \left(\! \begin{array}{c} 0\\0\\1 \end{array} \!\right) \ee^{-\chi \tau} + 
C_2 \left(\!\! \begin{array}{c} 1\\-1\\-1 \end{array} \!\!\right) \ee^{-\tau} - 
C_3 \left[\! \left(\!\! \begin{array}{c} 1\\-1\\-1 \end{array} \!\!\right)\tau + 
\left(\!\! \begin{array}{c} 0\\-1\\0 \end{array} \!\!\right) \!\right] \ee^{-\tau} . 
\eeq
The solution in this case corresponds to the scenario in which $\mu_2\rightarrow0$ 
as $\tau\rightarrow\tau_c<\infty$.  Let us assume 
\beq 
\mu_2 \sim \ol{\mu}_2 (\tau_c - \tau) , \quad N_2 \rightarrow \ol{N}_2 , \quad 
M_2 \rightarrow \ol{M}_2, \quad  {\rm as } \;\; \tau \rightarrow \tau_c, 
\eeq 
then $t_2 = \frac{2}{a}\int \frac{d\tau}{\mu_2(\tau)}$ implies 
$\tau_c-\tau=\ee^{-a\ol{\mu}_2t_2/2}$ so that $\tau\rightarrow\tau_c$ 
corresponds to $t_2\rightarrow\infty$. We thus have $\mu_2 \sim \ol{\mu}_2 
\ee^{-a\ol{\mu}_2t_2/2}$.  The solution of equation (\ref{logS2eq}) yields 
\beq 
S_2 \sim \ol{S}_2 (\tau_c-\tau)^{2k\ol{M}_2/a\ol{\mu}_2}  
=  \ol{S}_2 \ee^{-k\ol{M}_2t_2} . 
\eeq 

If $\half a\ol{\mu}_2 > k \ol{M}_2$, then this solution 
predicts that $\mu_2$ decays faster than $S_2$; however,  
towards the end of the second timescale 
the dominant terms in the equation for $\mu_2$ are 
$\mu'_2 = k S_2 M_2 - \half a \mu_2 N_2$.  
This equation implies that $\mu_2$ cannot decay faster than 
$S_2$, and thus we expect $\half a \ol{\mu}_2 \leq k \ol{M}_2$.  
If equality holds, then both $\mu$ and $S$ reach ${\cal O}(\ep^{1/5})$ 
at the same time.   In summary, towards the end of this timescale, 
we find $\mu_2$ and $S_2$ decaying exponentially in $t_2$ with 
\beq 
S_2 \sim \hat S_2 \ee^{-k\ol{M}_2 t_2} , \qquad 
\mu_2 \sim \ol{\mu}_2 \ee^{-a\ol{\mu}_2 t_2/2} , 
\eeq 
together with $N_2\rightarrow\ol{N}_2$ and $M_2\rightarrow\ol{M}_2$.  
This agrees with our observations of the numerical solution discussed 
in the opening paragraph of this subsection. The system then passes 
straight into Timescale III, which becomes relevant when $kSM\sim S_0$. 
This occurs when $S={\cal O}(\ep^{1/5})$ which happens when 
$t_2 \sim t_{2c} = (2/(5k\ol{M}_2)) \log(1/\ep)$. 

However, if $k\ol{M}_2>\half a \ol{\mu}_2$ then $S_2$ decays 
at a faster rate than $\mu_2$ and there is an further timescale between 
this and timescale III (below) over which $S$ saturates while $\mu$ 
decreases further. This timescale is given by 
\beq 
S=\ep^{1/5} \hat S , \quad \mu=\ep^\gamma \hat\mu , \quad 
N=\ep^{-1/5} \hat N , \quad M=\ep^{-1/5} \hat M , 
\eeq 
for some $-1/5 < \gamma < 1/5$ and 
$t=\ep^{-1/5}t_{1c}+\ep^{1/5}\log(1/\ep)t_{2c}+\ep^{1/5}\hat t$. 
The governing equations are then 
\beq 
\hat S' = S_0 - k \hat S \hat M , \quad 
\hat\mu' = -\half a \hat \mu \hat N , \quad 
\hat N' = 0 , \quad \hat M' = 0 , 
\eeq 
where prime denotes $d/d\hat t$. Over this timescale, 
$M$ and $N$ remain constant, $\hat S$ equilibrates to $S_0/k\hat M$ 
and $\hat\mu$ continues to decrease exponentially, 
with rate $-\half a \hat N$. When $\mu$ reaches ${\cal O}(\ep^{1/5})$, 
timescale III is entered; this occurs when $\hat t={\cal O}(\log(1/\ep))$, 
so the relationship between $t_3$ and $t$ in Timescale III given below 
remains valid following a redefinition of $t_{2c}$ to incorporate this 
extra shift in time. 

%------------
\subsubsection{Timescale III for smaller $\chi$: 
$t=\ep^{-1/5}t_{1c} + \ep^{1/5}\log(1/\ep)t_{2c} + {\cal O}(\ep^{1/5})$ }

In the new timescale we have 
\beq
S = \ep^{1/5} S_3 , \quad \mu = \ep^{1/5} \mu_3 , \quad 
N = \ep^{-1/5} N_3 , \quad M = \ep^{-1/5} M_3 , 
\lbl{t3sm-scal} \eeq
together with $t = t_{1c} \ep^{-1/5} + t_{2c} \ep^{1/5}\log(1/\ep) 
+ \ep^{1/5} t_3$.  Using prime to denote time derivative with 
respect to the new time variable, $t_3$, the leading order equations are 
\beq \begin{array}{rclcrcl}
S'_3 & = & S_0 - k S_3 M_3 , && N'_3 & = & 0 , \\ 
\mu'_3 & = & k S_3 M_3 - \half a (1+\chi) \mu_3 N_3, && M'_3 &=& 0 . 
\end{array} \eeq 
Over this rapid timescale, $M$ and $N$ do not change from their 
values at the end of the TII whilst $\mu$ and $S$ equilibrate to the 
values 
\beq
S_3 \rightarrow \frac{S_0}{kM_3} , \qquad 
\mu_3 \rightarrow \frac{2S_0}{a (1+\chi) N_3} , 
\quad {\rm as} \;\; t_3\rightarrow \infty .
\eeq
The evolution of $M$ and $N$ occurs over a longer timescale with 
$\mu$ and $S$ constrained to their respective local equilibrium values. 

%------------
\subsubsection{Timescale IV for smaller $\chi$: 
$t=\ep^{-1/5}t_{1c} + {\cal O}(\ep^{-1/5})$ }

Since all concentrations approached constants at the end of TIII, 
their magnitudes remain unchanged for the fourth timescale, 
only the scaling for $t$ changes, we now write 
\beq
S = \ep^{1/5} S_4 , \quad \mu = \ep^{1/5} \mu_4 , \quad 
N = \ep^{-1/5} N_4 , \quad M = \ep^{-1/5} M_4 , 
\eeq
with $t= t_{1c} \ep^{-1/5} + t_{2c} \ep^{1/5}\log(1/\ep) + \ep^{-1/5} t_4$, 
and so obtain the equations 
\beq \begin{array}{rclcrcl} 
0 & = & S_0 - k S_4 M_4 &&  
0 & =& k S_4 M_4 - \half a (1+\chi) \mu_4 N_4 \\ 
N'_4 & = & - \half a \chi \mu_4 N_4 && 
M'_4 & = & \half a \mu_4 ( N_4 - \chi M_4 ) . 
\end{array} \eeq 
These imply $kS_4M_4=S_0$ and $\mu_4N_4=2S_0/a(1+\chi)$, 
thus we have the solution 
\beq 
N_4 = \frac{\chi S_0 (t_{4c}-t_4)}{1+\chi} , \qquad
\mu_4 = \frac{2}{a\chi(t_{4c}-t_4)} , \qquad 
S_4 = \frac{S_0}{kM_4} , 
\lbl{t4sol} \eeq 
\beq 
M_4=C(t_{4c}-t_4)-\frac{S_0^2(t_{4c}-t_4)\log(t_{4c}-t_4)}{2(1+\chi)} , 
\lbl{t4M4sol} \eeq 
for some constant $C$.  This timescale ends with $\mu_4$ and 
$S_4$ increasing hence becoming larger in the next timescale, 
and $M_4$, $N_4$ decaying hence being smaller in the 
next timescale.  

%------------
\subsubsection{Timescale V for smaller $\chi$: 
$t=\ep^{-1/5}(t_{1c}+t_{4c}) + {\cal O}(1)$ }

The final timescale is given by all of $S$, $\mu$, $N$ and $M$ 
being ${\cal O}(1)$ and varying on an ${\cal O}(1)$ timescale, 
thus corresponds to the approach to the global equilibrium 
solution given by (\ref{ssss-mu-S})--(\ref{ssss-NM}). 

%------------
\subsubsection{Summary}

For all values of the parameters, the solution starts with a long 
induction period during which the concentration of the precursor 
species $S$ becomes large.  There follows a short period of  
very rapid kinetics where all concentrations become large, 
and then the precursor and monomer concentrations decay. 
For larger values of $\chi$ the concentration of polymers 
also decays and the steady-state is reached relatively rapidly. 
For smaller $\chi$ the monomer concentration and that of the 
source species become very small and the polymer concentrations 
(mass-weighted and number weighted) both remain high and 
slowly evolve to their steady-state values over a longer timescale, 
which is of similar length to the induction timescale. 
These two distinct behaviours are illustrated in Figures 
\ref{kin-big-fig} and \ref{kin-sm-fig}. 

%----------------------------
\subsection{Stability of the achiral growing solution}
\label{stab-sec}

Having determined the form of the kinetic behaviour for the 
achiral solution ($\eta=\theta=\de=0$), we now consider 
the linear stability of this solution. Assuming there is some 
small random perturbation during the evolution, we ask 
whether a perturbation grows or decays as time progresses.    
We shall use the already determined solution for $S(t)$, $\mu(t)$ 
$N(t)$ and $M(t)$ (from section \ref{grow-sec}), and assume 
that the perturbation does not make any alteration to these total 
concentrations of source, monomer and polymer at leading order. 
This assumption was seen to be valid in the case of 
the steady-state solution, and since the kinetic equations 
are symmetric under the transformation $(\de,\theta,\eta) 
\rightarrow (-\de,-\theta,-\eta)$, we expect modifications 
to $\mu,N,M,S$ also to be of second order in $\de,\theta,\eta$. 
{}From equations (\ref{kine})--(\ref{kinde}) we have 
\beq
\left( \begin{array}{c} \dot\eta\\[1ex] \dot \theta \\[1ex] \dot\delta \end{array} \right) 
= \half a\mu \left( \begin{array}{ccc} - \frac{2\mu}{M}\!-\!\frac{N}{M} & 
\frac{N}{M} & \chi\!+\!\frac{4\mu}{N}\!+\!\frac{N}{M} \\[1ex] 
0 & - \frac{\mu}{N} & \chi\!+\!\frac{2\mu}{N} \\[1ex] 
\frac{2kfSM}{a\mu^2} & \frac{(\chi-1)N}{\mu} & -\!2\!-\!\frac{2kSM}{a\mu^2}
\end{array}\right)\left(\begin{array}{c}\eta\\[1ex] \theta\\[1ex] \delta\end{array}\right).
\lbl{glob-stab}\eeq

%----------------------------
\subsubsection{Stability of achiral solution in Timescale I}

We focus our attention on the linear stability of $\delta(t)$, 
$\theta(t)$ and $\eta(t)$ as given by (\ref{lin-eta})--(\ref{lin-de}).  
In the first timescale, using the scalings (\ref{T1-scal}), 
we find the simplified linear stability problem
\beq
\frac{d}{dt_1} \left( \begin{array}{c} \eta \\[1ex] \theta \\[1ex] \de 
\end{array}\right) = \half a \mu_1^2 \left( \begin{array}{ccc} 
-\frac{2}{M_1} & 0 & \frac{4}{M_1} \\[1ex] 
0 & -\frac{1}{N_1} & \frac{2}{N_1} \\[1ex] 
\frac{2kS_1M_1f}{a\mu_1^3} & 0 & -\frac{2kS_1M_1}{a\mu_1^3} 
\end{array}\right)\left(\begin{array}{c}\eta\\[1ex] \theta\\[1ex] \de\end{array}\right).
\eeq 
It is clear that this matrix has a simpler structure than the general case, which 
has only one zero entry, as given in equation  (\ref{stab-mat}).  The eigenvalues 
of the matrix above are all negative if $f<1/2$, but if $f>1/2$ then one is positive, 
indicating that a perturbation away from $\theta,\eta,\de=0$ would increase in 
size as time progresses. The temporal evolution of such perturbations is  
non-trivial since $M_1$, $N_1$, $\mu_1$ and $S_1$ are all time-dependent. 

%----------------------------
\subsubsection{Stability of achiral solution in Timescale II, larger $\chi$}

{}From equation (\ref{glob-stab}) for the linear stability of the 
achiral solution to chiral perturbations,  and (\ref{t2-bigchi-sol}) 
for the concentrations $S_2$, $\mu_2$, $N_2$ and $M_2$, we obtain 
\beq 
\frac{d}{dt_2}\left(\begin{array}{c}\eta\\ \theta\\ \delta\end{array}\right)=
\frac{a\chi }{4 k t_2} \left( \begin{array}{ccc} -\chi & \half & 5\chi \\ 
0 & -\chi & 3\chi \\ 3f & 1 & -5 \end{array} \right) 
\left( \begin{array}{c} \eta \\ \theta \\ \delta \end{array} \right) , 
\eeq 
Stability is determined by the eigenvalues of the matrix, 
which satisfy the cubic equation 
\beq
0 = \lambda^3 + (2\chi+5)\lambda^2 + (\chi^2+7\chi+15f\chi)\lambda 
+ 2\chi^2 - \half\chi f (9+30\chi)  . 
\eeq
Applying the Routh Hourwitz criteria (see the end of Section \ref{sec31} 
for details),  we find that an instability occurs if $f>4\chi/(9+30\chi) \sim 2/15$ 
(for large $\chi$).  

%----------------------------
\subsubsection{Stability of achiral solution in Timescale II, smaller $\chi$}

At the end of Timescale II if $\chi$ is small then the matrix in equation 
(\ref{glob-stab}) has the form 
\beq 
\frac{d}{dt_2}\left(\begin{array}{c}\eta\\[1ex]\theta\\[1ex]\delta\end{array}\right)=
\half a\mu_2 \left(\begin{array}{ccc}  -\frac{\ol{N}_2}{\ol{M}_2} 
& \frac{\ol{N}_2}{\ol{M}_2} & \chi+\frac{\ol{N}_2}{\ol{M}_2}\\ 0&0&\chi\\
\frac{2kfS_2M_2}{a\mu_2^2} & \frac{(\chi-1)\ol{N}_2}{\mu_2} & 
-\frac{2kS_2M_2}{a\mu_2^2}  \end{array}\right) 
\left( \begin{array}{c} \eta \\[1ex] \theta \\[1ex] \delta \end{array} \right) . 
\eeq 
All elements in the bottom row of this matrix are divergent since 
$\mu_2,S_2\rightarrow0$ as $t_2\rightarrow\infty$.  When $t_2$ becomes 
large the eigenvalues of this matrix are given by solutions of the cubic 
\beqa 
0 & = & \lambda^3 + \frac{2kS_2\ol{M}_2}{a\mu_2^2} \lambda^2  
+ \frac{2kS_2\ol{M}_2}{a\mu_2^2} \left[ \frac{N_2}{M_2}  + 
\frac{a\chi\mu_2\ol{N}_2}{2kS_2\ol{M}_2} 
- f \left(\chi+\frac{\ol{N}_2}{\ol{M}_2}\right)  \right] \lambda \nn \\ && 
- \left( \frac{2 k f S_2 \chi \ol{N}_2}{a \mu_2^2} + 
\frac{\chi(\chi-1)\ol{N}_2^2}{\ol{M}_2\mu_2} \right) . 
\eeqa 
Applying the Routh-Hourwitz criteria for the signs of the real parts of the 
eigenvalues, we find stability of the achiral solution requires $A>0$ 
(which always holds), 
and $C>0$  which fails when $f>a \ol{N}_2 \mu_2 / 2k\ol{M}_2 S_2$; 
and a similar but more stringent inequality from the condition $AB>C$.  
So if $S_2$ decays faster than $\mu_2$ then the symmetric solution 
is stable, and if $S_2$ and $\mu_2$ decay at the same rate the 
stability depends on $\ol{N}_2$ and $\ol{M}_2$ and requires 
$f> 1/(2(1-k\ol{M}_2/a\ol{N}_2))$.  

%----------------------------
\subsubsection{Stability of achiral solution in Timescale III, smaller $\chi$}

With the scalings of (\ref{t3sm-scal}) the matrix (\ref{glob-stab}) takes the form 
\beq 
\frac{d}{dt_3} \left( \begin{array}{c} \eta\\[1ex] \theta\\[1ex] \delta 
\end{array} \right) = \half a \mu_3 \left( \begin{array}{ccc} 
-\frac{N_3}{M_3}\ep^{2/5}  & \frac{N_3}{M_3}\ep^{2/5}  & 
\left(\chi\!+\!\frac{N_3}{M_3}\right) \ep^{2/5}  \\[1ex] 
0 & 0 & \chi\ep^{2/5} \\[1ex] \frac{2kfS_3M_3}{a\mu_3^2} 
& - \frac{N_3}{\mu_3} & - \frac{2kS_3M_3}{a\mu_3^2} 
\end{array} \right) \left( \begin{array}{c} \eta\\[1ex] \theta\\[1ex] \delta 
\end{array} \right) . 
\lbl{t3sm-stab}\eeq 
Note that the final row of entries are asymptotically larger than the other entries. 
Taking only the leading order entries for each term, the cubic governing stability 
can be written  
\beq
0 = \frac{a\mu_3^2\ep^{2/5}}{2kS_3M_3} \lambda^3 + \lambda^2 + 
\left( \frac{N_3}{M_3} - f \left( \chi+\frac{N_3}{M_3} \right) \right) \lambda + 
\frac{ a \chi\mu_3 N_3^2}{2 k S_3 M_3^2} - \frac{f\chi N_3}{M_3} . 
\eeq
Rewriting this as $\lambda^3 + A \lambda^2 + B \lambda + C$, the 
Routh-Hourwitz criterion $A>0$ is always satisfied. The condition 
$C>0$ implies $f<a\mu_3 N_3 / 2 k S_3 M_3$ which, at large times, reduces to 
$f<1$, and as such is met in the large-time limit. Finally, $AB>C$ fails if 
$f>N_3/(N_3+\chi M_3)$ indicating an instability; whilst this depends on 
the unknown number-weighted and mass-weighted polymer concentrations, 
an instability is certainly possible for large enough values of the
fidelity parameter, $f$. 

%----------------------------
\subsubsection{Stability of achiral solution in Timescale IV, smaller $\chi$}

Since the scalings for the concentrations $S$, $\mu$, $N$ and $M$ 
are as in timescale III, the only change from (\ref{t3sm-stab}) is in the rate 
of growth of the perturbations since $d/dt_4$ and $d/dt_3$ scales differently 
with time and $N_4,M_4$ are now time-dependent; we now have 
\beq 
\frac{d}{dt_4} \left( \begin{array}{c} \eta\\[1ex] \theta\\[1ex] \delta 
\end{array} \right) = \half a \mu_4  \left( \begin{array}{ccc} 
-\frac{N_4}{M_4} & \frac{N_4}{M_4} & \chi\!+\!\frac{N_4}{M_4}\\[1ex] 
0 & 0 & \chi\\[1ex] \frac{2kfS_4M_4\ep^{-2/5}}{a\mu_4^2} 
& - \frac{N_4\ep^{-2/5}}{\mu_4} & - \frac{2kS_4M_4\ep^{-2/5}}{a\mu_4^2} 
\end{array} \right) \left( \begin{array}{c} \eta\\[1ex] \theta\\[1ex] \delta 
\end{array} \right) . 
\eeq 
As the solution progresses through timescale IV, the criterion for an instability 
to exist, namely $f>N_4/(N_4+\chi M_4)$, becomes easier to satisfy, since 
$N_4$ decays slightly faster than $M_4$, as shown by equations 
(\ref{t4sol})--(\ref{t4M4sol}).   

%-----------------------
\subsection{Summary}

In this section we have analysed the kinetics of the concentrations of monomer, 
source and polymer as they evolve from zero to steady-state following a 
symmetric (achiral) solution. 

For larger values of $\chi$ we have found three regimes through which 
the system evolves.  Firstly there is a long induction period 
(of ${\cal O}(\ep^{-1/5})$) during which the source material  
builds up (until $S$ becomes ${\cal O}(\ep^{-1/5})$); during this 
time the monomer concentrations remain small (${\cal O}(\ep^{3/5})$) 
and the concentrations of polymer are extremely small (${\cal O}(\ep)$). 
This timescale ends abruptly as the catalytic feedback of polymer 
accelerates the breakdown of $S$ into monomer. 
In the second timescale, which is very brief (${\cal O}(\ep^{1/5})$), 
all monomer and polymer concentrations become large 
(${\cal O}(\ep^{-1/5})$) and decay towards steady-state. Finally, over the 
third timescale all concentrations converge to their steady-state values. 

In section \ref{stab-sec} we analysed the linear stability of the 
growing achiral state through the sequence of timescales.  
By linear stability we mean that small external random forces which 
cause a chiral imbalance will be damped and reduce in amplitude.  
An instability indicates that such a perturbation will grow and so the system 
will undergo a symmetry-breaking bifurcation to occur during the evolution. 

As the system approaches steady-state in the third timescale we expect to regain 
the stability criteria $f>f_c(\chi)$ with $f_c$ given by equation (\ref{fcdef}). 
However, this criterion for the instability of the achiral solution is not valid in all 
the timescales; in the first timescale we find the alternative criteria of $f>\half$ 
for symmetry-breaking to occur.   Note that this is independent of $\chi$, since 
in the first timescale the polymer and monomer concentrations are so low that 
inhibition of a homochiral polymer by a monomer of the opposite chirality is 
negligible; this leads to a considerable simplification of the linear stability analysis 
during the first timescale. In the second timescale we find an instability for $f<2/3$. 

For $f<2/9$ the achiral solution is always stable to such perturbations; 
whereas for $2/9<f<1/2$ the system is unstable to such perturbations 
only in the final stage of the kinetics, when the equilibrium solution is 
being approached and even then only for some values of $\chi$.  
For $1/2<f<2/3$ the kinetics are more complicated 
since the system is unstable during the first timescale, but linearly stable 
during the second timescale and unstable during the final approach to 
equilibrium.  Thus for these $f$-values the chiral purity may oscillate, but 
the system will eventually approach a chiral state. 

For smaller values of $\chi$ the kinetics are more complex: there is still 
a long induction time followed by a period of rapid kinetics. This is more 
complicated, being split into various timescales; however the end result 
is always low concentrations of monomer and source species and large 
concentrations of polymer.  There follows a long timescale over which 
the polymer concentrations reduce towards steady-state and the 
monomer concentrations and source species increase to steady-state.  

For smaller values of $\chi$ an instability occurs in Timescale I for $f>\half$, 
the instability persists in timescale II,  now depending on 
$f> 1/(2(1-k\ol{M}_2/a\ol{N}_2))$.  In timescale III and IV an instability 
requires $f> N/(N+\chi M)$, and  at steady-state $f>4/5 - 34\chi/25$. 
Thus, for smaller $\chi$ there is still the possibility of a symmetry-breaking 
bifurcation occurring, though it requires a larger fidelity parameter.  

Although there advantages in $\chi$ being large if one is seeking a 
symmetry-breaking bifurcation, it should be noted that smaller $\chi$ has 
other advantages, in that it allows polymers to form in larger concentrations, 
and these persist for longer times. 

%----------------------------------------------------------------------
\section{Perfect fidelity}
\label{fe1-sec}

In the extreme case $f=1$ the feedback mechanism 
breaks down the precursor species ($S$) into chirally pure 
monomers with unit probability. Instead of a chirally 
pure homochiral steady state, there is a bifurcation to a state in 
which $\delta$ asymptotically approaches $\pm1$. 
We refer to such a state as a fully-bifurcated state. From 
equation (\ref{fde-def}) we see that such a state cannot  
arise if $f<1$. The large-time asymptotics of the 
fully-bifurcated state differ significantly from $f<1$ since 
now there is no steady-state solution; instead we find 
unlimited growth of one set of homochiral polymer sequences 
and decay to zero for the sequences of opposite handedness.   
In this case (still ignoring $\ep$) the large time asymptotics 
are given by 
\beq
M=\hat M t , \quad N=\hat N t^{1/3} , \quad 
\mu = \hat \mu t^{-1/3} , \quad S=\hat S t^{-1} , 
\eeq
\beq
\de=1-\tilde\de,\qquad\theta=1-\tilde\theta,\qquad \eta=1-\tilde\eta.
\eeq

Assuming $\tilde\delta$, $\tilde\theta$ and $\tilde\eta$ all decay 
to zero in the large time limit, we find the leading order equations 
\beq
S_0 = k \hat S \hat M , \qquad 
\hat M = a \hat \mu \hat N , \qquad 
\rec{3} \hat N = a \hat\mu^2
 S_0 = a\hat\mu\hat N , 
\eeq
\beq
\dot{\tilde{\de}} = - a \hat N (\chi-1) t^{1/3} \tilde \theta ,\qquad 
\dot{\tilde{\theta}} = - a \chi \hat\mu t^{-1/3} \tilde \theta , \qquad 
\dot{\tilde{\eta}} = - a \chi \hat\mu t^{-1/3} \tilde\eta . 
\eeq
These imply 
\beq
\hat S = \frac{1}{k} , \qquad
\hat M = S_0 ,\qquad
\hat N = \frac{S_0^{2/3}}{a^{1/3}} ,\qquad 
\hat \mu = \frac{S_0^{1/3}}{a^{2/3}} . 
\eeq
\beq
\tilde\de \sim \exp\left(-\mfrac{3}{4} a \hat N(\chi-1)t^{4/3}\right) ,
\qquad
\tilde\theta , \tilde\eta \sim 
\exp\left(-\mfrac{3}{2} a\chi\hat\mu t^{2/3} \right) . 
\eeq
Thus in terms of the concentrations of each chirality we have 
\beq
R_1 \sim \mu \sim \left( \frac{S_0}{a^2t} \right)^{1/3} , \qquad 
R \sim N \sim \left( \frac{S_0^2 t}{a} \right)^{1/3} , \qquad 
P \sim M \sim S_0 t ,
\eeq
together with $S \sim 1/kt$, and 
\beqa
L_1 &\sim&\tilde{L}_1t^{-1/3}\exp(-\mfrac{3}{4} (\chi-1)(aS_0t^2)^{2/3}),\\
\{L,Q\}&\sim&\{\tilde{L},\tilde{Q}\}t^{1/3}\exp(-\mfrac{3}{2}\chi(aS_0t^2)^{1/3}),
\eeqa
for some constants $\tilde{L}_1$, $\tilde{L}$, $\tilde{Q}$.

So in this case no finite steady-state solution is approached. 
As one might expect from the asymptotic expansions 
(\ref{Rnuasy})--(\ref{Lnuasy}) in the case $\de\rightarrow\pm1$ 
we observe the unbounded growth of one type of homochiral 
sequence and, specifically, unbounded growth in 
the number of chains ($N$), the mass of material in polymeric form 
($M$) and the average length ($M/N\sim (aS_0t^2)^{1/3}$). 
Concentrations of the sequences of opposite homochirality 
decay rapidly, and we expect that the average 
chain length approaches two, implying $\tilde{Q}=2\tilde{L}$.

All the above analysis has been for the simplified case for which $\ep=0$; 
however, with one chain type decaying to arbitrarily small 
concentrations we may expect that the ${\cal O}(\ep)$ term 
in (\ref{L1doteq}) is no longer negligible in this limit.  Retaining 
the ${\cal O}(\ep)$ term in this equation yields a slightly different 
scaling in the large time asymptotics for the high-fidelity case $f=1$. 
We now have 
\beq
R_1 \sim \left( \frac{S_0}{3a^2t} \right)^{1/3} , \qquad 
R \sim \left( \frac{3S_0^2 t}{a} \right)^{1/3} , \qquad 
P \sim S_0 t , 
\eeq
together with $S_0\sim1/kt$ and 
\beq
L_1 \sim \frac{\ep}{k\chi (3a^2S_0^2t^4)^{1/3}} , \qquad 
L \sim \frac{\ep^2}{k^2\chi^3S_0 (3a^2S_0^2)^{1/3} t^3} ,  
\eeq
\beq
Q \sim \frac{2\ep^2}{k^2\chi^2(3a^2S_0^2)^{1/3} t^3} . 
\eeq
Thus once again we see the less common homochiral polymer 
sequences assuming concentrations which decay to zero, albeit 
now with the simpler form of algebraic decay,  
the typical polymeric length again asymptoting to two ($Q/L$). 
The dominant homochiral sequences grow in number, mass and 
average length (with $P/R\sim(a S_0 t^2/3)^{1/3}$). 
As $t\rightarrow\infty$ the chiral purity of the system 
approaches unity according to 
\beq 
\de\sim1-\frac{2\ep}{k\chi S_0 t},
\eeq 
\beq 
\theta\sim1-\frac{2\ep^2}{k^2\chi^3 S_0^2 (9aS_0t^{10})^{1/3}},\;\;
\eta\sim1-\frac{4\ep^2}{k^2\chi^3 S_0^2 (3a^2S_0^2)^{1/3} t^4} . 
\eeq 
Thus, we see the convergence to full chiral purity 
is more rapid for polymers than for monomers. 

In all the above analysis there is another chiral solution 
in which the left-handed homochiral polymer sequences 
are dominant, and the right-handed homochiral sequences 
have concentrations which decay to zero asymptotically. 

%----------------------------------------------
\section{Discussion}
\label{disc-sec}

After introducing our model in Section \ref{mod-sec}, we analysed 
its steady-states, and found that the symmetric steady-state solution 
exists for all parameter values, but that there are other solutions 
when the relative inhibition rate $\chi$ and the fidelity $f$ 
are large enough. The critical combination is 
\beq 
f > f_c = \frac{(4+2\chi)(1+2\chi)}{(5+6\chi)(1+3\chi)} . 
\eeq 
At the point $f=f_c$ there is a supercritical pitchfork bifurcation, 
where two unstable steady-state solutions connect to the solution 
$\de=0$ and make $\de=0$ unstable for $f>f_c$.  When this 
inequality is satisfied there are two stable steady-state 
solutions with $\de$ non-zero, that is there are chiral solutions 
as well as the achiral solution, and the achiral solution is unstable, 
so any physical system will generically be attracted to one or other 
of the chiral solutions. An important effect to note from 
this formula is the role that cross-inhibition plays in making the 
asymmetric solutions accessible at low values of the fidelity. 
For small $\chi$, the fidelity has to exceed $f_c=0.8$ in order 
to obtain a symmetry-breaking solution; whereas at large $\chi$, 
this bifurcation point reduces to $f_c=0.22$ -- a dramatic reduction. 

In Section \ref{kin-sec} we analysed the kinetics of 
chain growth in a symmetric system, and found that 
there is a long induction time, during which a large 
stock of precursor chemical accumulates; 
an approximate, linear stability calculation shows that 
during this time, the achiral solution is unstable if $f>1/2$. 
This behaviour is followed by a short timescale over which 
the precursor species is converted to monomers which are 
then polymerised.  For this short time, monomers and chains 
are present in large concentrations. The concentrations 
of chains, monomers and precursor then all decay to 
their steady-state values, which, if the parameters 
$f,\chi$ are such that an asymmetric steady-state exists, 
and the system has experienced some external perturbation 
away from the symmetric state, will be the chiral
steady-state discussed earlier (section \ref{asym-sss-sec}). 
In such a state {\em both} 
monomers and chains have a net chirality or handedness.  
Even for quite modest values of the chiral purity of monomer 
(say $\de=0.7$), the chiral purity of chains is extremely close 
to unity ($\theta=0.990$, $\eta=0.995$ at $\chi=2$); see 
Figure \ref{th-et-fig}, and compare Figures \ref{bif-fig} 
and \ref{bif-fig2}.

Finally we have described the large-time asymptotics 
of the `fully' bifurcated case which arises when 
$f=1$, wherein chiral purities ($\de$, $\theta$ and $\eta$) 
approach unity in the large time limit; this remains true even 
when we reintroduce the term which describes the 
slow spontaneous achiral decay of precursor 
species into both enantiomeric forms of monomer. 

%-------------------------------------------------------
\section{Conclusions}
\label{conc-sec}

In previous work we showed how qualitatively similar instabilities 
can lead to the massive amplification of self-replicating RNA polymer 
sequences over less efficient replicators (Wattis \& Coveney, 1999). 

In the present paper we have shown that an initially achiral system capable of 
stepwise polymerisation to homochiral polymer sequences with inhibition 
from the opposite-handed monomer, is subject to strong instabilities 
that drive the system overwhelmingly to one or other handedness for 
all homochiral sequences present.  Mechanisms within this class may 
have played a r\^{o}le during the early stages of molecular evolution 
in determining the chirality of biologically relevant macromolecules, 
such as nucleotides and proteins, and there is experimental evidence 
of this behaviour in the literature, for examples see Hitz \etal\ 
(2001, 2002, 2003) and Joshi (2000). 
Although in the system studied by Joshi, addition of the correct enantiomer 
to a growing polymer chain is more favourable than the wrong one, 
we have shown that if this cross-inhibition is stronger, then the system is 
more likely to undergo a symmetry-breaking bifurcation. 

Studies of this kind confirm the scope and power of modern methods of 
theoretical analysis for nonlinear dynamical systems of the kind that abound 
along the pathway towards the origins of life (Coveney, 1994). 

%------------------------
\subsection*{Acknowledgements}

We are grateful to Prof Pat Sandars for useful conversations about our 
models, which are based on his work.  JADW would like to thank Fernando 
da Costa for conversations about this subject, and the British Council for 
funding a visit to Institute Superior Tecnico, Lisbon, Portugal where 
some of the work was carried out.  PVC is grateful to EPSRC for funding 
under EPSRC Reality Grid grant GR/R67699. 

%-------------------------------------------------------

\bibliographystyle{plain}

\end{document}